\documentclass{aa}
\usepackage{graphicx}
\usepackage{txfonts}
\begin{document}
  \title{Gas stripping in galaxy clusters: a new SPH simulation 
approach}

    \author{
        P. J\'achym     \inst{1}
    \and
        J. Palou\v s    \inst{1}
    \and
        J. K\"oppen     \inst{1,2,3,4}
    \and
        F. Combes       \inst{5}
           }
    \offprints{P. J\'achym }

    \institute{Astronomical Institute,
           Academy of Sciences of the Czech Republic,
           Bo\v cn\' \i \ II 1401, 141 31 Prague 4, Czech Republic\\
           \email{jachym@ig.cas.cz, palous@ig.cas.cz}
          \and
            Observatoire Astronomique de Strasbourg,
            11 Rue de l'Universit\'e,
            F--67000 Strasbourg, France\\
            \email{koppen@astro.u-strasbg.fr}
         \and
            International Space University,
            Parc d'Innovation,
            1 Rue Jean-Dominique Cassini,
            F--67400 Illkirch-Graffenstaden, France
         \and
            Institut f\"ur Theoretische Physik und Astrophysik,
            Universit\"at Kiel,
            D--24098 Kiel, Germany
          \and
            Observatoire de Paris, LERMA, 61 Av. de l'Observatoire,
            75014, Paris, France
              }

   \date{Received September 25, 2006; accepted May 22, 2007}

\titlerunning{Gas stripping in galaxy clusters: a new SPH simulation 
approach}
\authorrunning{J\'achym, Palou\v s, K\"oppen \& Combes}

\abstract
{}
{The influence of a time-varying ram pressure on spiral galaxies in 
clusters is explored with a new simulation method based on the N-body 
SPH/tree code GADGET.}
{We have adapted the code to describe the interaction of two different
gas phases, the diffuse hot intracluster medium (ICM) and the denser
and colder interstellar medium (ISM). Both the ICM and ISM components
are introduced as SPH particles. As a galaxy arrives on a highly radial
orbit from outskirts to cluster center, it crosses the ICM density peak
and experiences a time-varying wind.} {Depending on the duration and
intensity of the ISM--ICM interaction, early and late type galaxies in
galaxy clusters with either a large or small ICM distribution are found
to show different stripping efficiencies, amounts of reaccretion of the
extra-planar ISM, and final masses. We compare the numerical results
with analytical approximations of different complexity and indicate the
limits of the Gunn \& Gott simple stripping formula.} {Our 
investigations
emphasize the role of the galactic orbital history to the stripping
amount. We discuss the contribution of ram pressure stripping to the
origin of the ICM and its metallicity. We propose gas accumulations
like tails, filaments, or ripples to be responsible for stripping in
regions with low overall ICM occurrence.}


\maketitle
%________________________________________________________________

\section{Introduction}
Galaxies in clusters and rich environments are observed to be stripped 
of their interstellar medium, which quenches subsequent star formation. 
The stripping can be due either to tidal interactions, spiral galaxy 
mergers into ellipticals, or to ram-pressure stripping from the 
intracluster gas. Both tidal and ram-pressure stripping are always 
tightly linked. Gunn \& Gott (1972) assume that, after the formation of 
a galaxy cluster, the remaining gaseous debris is thermalized via shock 
heating to virial temperatures corresponding to random motions in the 
cluster, e.g. to a few times 10$^7$K. This hot plasma at densities of a 
few times 10$^{-3}$ cm$^{-3}$ influences the ISM in disks of spiral 
galaxies and can remove part of their gas through the ram pressure 
induces by the galaxy motions through the ICM.

The ram pressure stripping is difficult to model, since several complex 
gaseous phenomena are involved. The first aspect is the simple pressure 
force, and this can be modeled with a simple algorithm using ballistic 
and sticky particles to represent the gas (e.g. Vollmer et al.~2001). 
But the full hydrodynamical processes include thermal evaporation, 
turbulent and viscous stripping, and also outflows due to star 
formation.

The first notice of the 3D N-body/SPH simulations with the gravity tree 
to mimic the dynamical effect of the ram pressure on galaxies in 
clusters is by Kundi\' c et al. (1993). The same approach has been 
adopted by Abadi et al. (1999). The ICM is represented as a flow of 
particles along a cylinder of radius 30 kpc and thickness 10 kpc. A 
spiral galaxy is in face-on, edge-on, or an inclined orientation 
relative to flow of ICM particles. With the simulations they examine 
the radius up to which the ISM is removed and compare the results with 
the prediction of Gunn \& Gott (1972) that the ISM is removed from the 
disk if the ICM ram pressure exceeds the restoring force:
\begin{equation}\label{striprad}
\rho_{\rm ICM} v^2 \ge {\partial \Phi(r,z) \over \partial z} \Big|_{\rm 
max} \Sigma_{\rm ISM},
\end{equation}
where $\rho_{\rm ICM}$ is the ICM density, $v$ the relative velocity of 
the galaxy and the ICM, $\Phi(r,z)$ the total gravitational potential 
of the galaxy as a function of the galactocentric cylindrical 
coordinates, and $\Sigma_{\rm ISM}$ the ISM surface density. The 
conclusion is that Eq.~\ref{striprad} applies in the face-on galaxy 
orientation when the bulge does not dominate the disk gravity. In other 
cases, edge-on orientation and in central parts of galaxies, where 
bulge dominates, the stripping is less efficient. The time scale for 
gas removal is $\sim 10^8$ years.

The ISM is far from continuous, but instead consists of several 
components: hot, warm, and cold medium. Hot ionized and warm neutral HI 
medium are much more continuous compared to cloudy, cold, and molecular
component. The SPH description relates more to the continuous ISM 
components, while molecular clouds are not represented well in the SPH 
simulations. Vollmer et al.~(2001) adopt another approach, using N-body 
simulations with sticky particles representing the inelastic collisions 
between molecular clouds. The ram pressure is included with an 
additional friction force acting on clouds in the wind direction. 
Vollmer et al.~(2001, 2006) introduce a time dependent ram pressure, which 
corresponds to the variations of the ICM density and of relative 
ICM--ISM velocity along the galaxy orbit in the cluster. They allow a 
moving ICM, which may increase its velocity relative the ISM to values 
higher than 4000 km s$^{-1}$.

With an isothermal SPH gas model, Abadi et al.~(1999) show that gas 
stripping can be quite efficient in the core of rich clusters on a 
time-scale of 10$^7$ yrs. The most efficient is face-on orientation.
With a Eulerian code, Quilis et al.~(2000) probed the efficiency of
viscous coupling, enhanced by the presence of HI deficiencies in the 
center of galaxies. The ram pressure stripping does not strongly depend 
on the vertical structure and thickness of the gas disk (Roediger \&
Hensler 2005). In more detail, the effect of inclination of the moving
galaxy is important as long as the ram pressure is comparable to the
gas pressure in the galaxy plane. In general the effect is similar for
all inclinations except for edge-on (Roediger \& Br\"uggen 2006). The
orientation of the gas tail behind the galaxy is not a good tracer of
the galaxy motion on its orbit. Acreman et al. (2003) simulate the
infall of an elliptical galaxy and its gas stripping by the ICM showing 
that the formation of an X-ray wake at the first passage should be 
observable.

The principal competing mechanism for perturbing galaxies in clusters 
and stripping their gas is tidal interactions. High-speed galaxy 
encounters in clusters, galaxy harassment events, are efficient 
mechanisms for disturbing galaxy disks, and providing stars to the 
intracluster space (Moore et al.~1996). Also low-speed galaxy 
encounters in groups before the formation of the cluster are able to 
drag tidal stellar tails out and lead to galaxy mergers and formation 
of ellipticals, which are then known to be more numerous in clusters. 
The evolution of encounter debris in galaxy haloes has been followed by 
Mihos (2004).

Intracluster light is a consequence of the tidal stripping of galaxies. 
It has been determined that intra-cluster stars are older than galaxy 
stellar populations and more centrally concentrated in the cluster. The 
fraction of stars in the ICM is increasing with the richness of the 
cluster and it is usually a few percent of all stars, up to 20\% 
(Arnaboldi et al.~2003, Murante et al.~2004). If the ICM metallicity 
can be explained partly by gas stripping from galaxies (Domainko et 
al.~2006), it can also come from intra-cluster supernovae exploding in 
the ICM. The irregular structures in galaxy clusters found in X-ray are 
tracing cold fronts and shocks due to galaxy interactions or to the 
infall of a galaxy group.

The fate of the stripped ISM in clusters is not clear, because it can 
end very hot and become part of the ICM or stay quite cold condensing
into clouds with a molecular core and an HI outer boundary. Oosterloo
\& van Gorkom (2005) observed a large HI cloud near the center of the
Virgo cluster and suggest that this cloud has been ram-pressure
stripped from the galaxy NGC 4388. The dense clumps in the HI plume
might be molecular, and star formation could occur precisely at these
high-density places.

In this paper, we focus on computations of the ram pressure stripping,
to evaluate its efficiency more precisely and to follow the fate of 
the stripped gas. The particular goal is to take the finite time for 
this stripping interaction into account, as the galaxy passes quickly 
through the central region of the cluster. Most computations until now 
were carried out in the hypothesis of a stationary wind of hot gas. But 
the ram pressure acts only in a short lapse of time, as simulated by 
Vollmer et al.~(2001). After this short period, a certain fraction of 
the gas falls back onto the galaxy. These latter simulations did not 
take into account both the pressure forces and the hydrodynamical 
physics of the ISM--ICM interactions. Roediger, Br\"uggen \& Hoeft 
(2006) have followed the wake of gas produced by a quasi-stationary 
wind; here, on the contrary, we emphasize the impulsive character of 
ram pressure stripping.

Assuming a static ICM described with a $\beta$-profile, we explore the 
ISM--ICM interaction in detail during the galaxy crossing of the 
central part of the cluster. The mass-loss rates are examined as a 
function of the galaxy type, size, and mass. We are able to follow the 
gas stripped from the galaxy, forming a giant tail of material. We 
tackle the issue of the origin of the ICM and the actual role of gas 
stripping in building it as a function of the cluster mass and 
richness. The importance of gas stripping could account for the metal 
enrichment of the ICM as a function of cluster type (e.g. Domainko et 
al.~2006, Schindler et al.~2005). After presenting the model of 
stripped galaxies and of the cluster in sections 2 and 3, we introduce 
the initial conditions in section 4 and describe our simulation method 
in section 5. In section 6, we discuss the results of the numerical 
simulations, the simulation test is shown in section 7, and the 
equation of motion of an individual ISM gas element and the impulse 
approximation are given and applied to stripping in section 8. The 
present simulations are compared to other simulations in section 9. 
Section 10 presents the discussion and conclusion.

\begin{figure*}[ht]
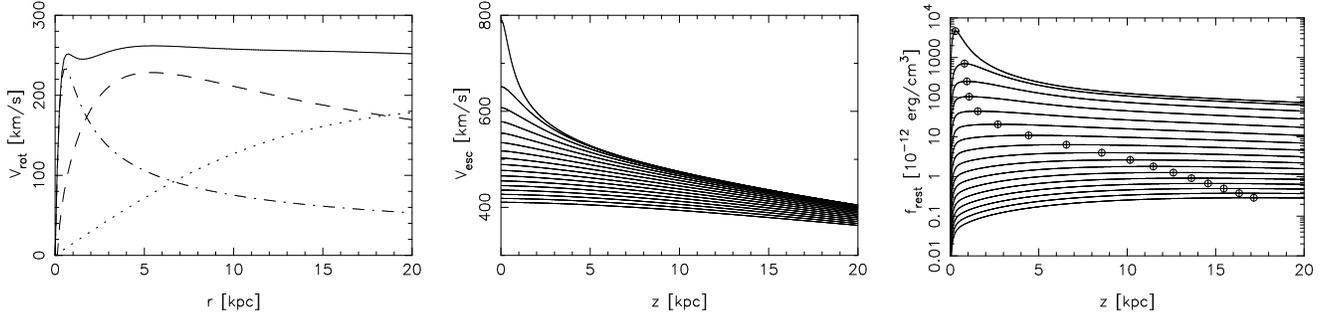

\centering
\includegraphics[width=0.3\textwidth,angle=0]{rotcurve.ps}
\hspace{0.25cm}
\includegraphics[width=0.3\textwidth,angle=0]{vesc.ps}
\hspace{0.25cm}
\includegraphics[width=0.3\textwidth,angle=0]{frest_z.ps}
\caption{\small LM model galaxy. Left: Rotation curve with
contributions of the halo (dotted), bulge (dash-dotted), and disk
(dashed) components. Center: Escape velocity $v_{esc}$ as a
function of the $z$-distance behind the disk plane for 16 radii
$r$ starting with $r=1$ kpc (top curve) and a spacing of 1 kpc.
Right: Gravitational restoring force $f_{rest}$ as a function of
$z$ for 16 values of radius $r$ from 1 kpc (top curve) to 16 kpc. 
Maxima in $z$ for individual radii are denoted with circles.
}\label{frest}
\end{figure*}

\section{The galaxy model}\label{sec:galaxy}
Our model of a spiral galaxy is a standard three-component halo + bulge 
+ disk configuration. The halo and bulge components are given with 
spherically symmetric, Plummer density distributions
\begin{equation}\label{plummer}
\rho_{b,h}(d) = \rho_{0}\,\frac{a_{b,h}^{5}}{(d^2+
a_{b,h}^2)^{5/2}},
\end{equation}
where $\rho_{0}=3M_{b,h}/4\pi a_{b,h}^3$ is the central density, $d$ 
the distance from the galactic center, $M_b, M_h$ are total masses, and 
$a_b$ and $a_h$ are the radial scaling factors corresponding to the 
bulge or halo.

The axially symmetric disk follows an infinitely thin Toomre-Kuzmin 
disk multiplied by a sech$^2 (z/z_0)$ term that defines its isothermal
vertical profile,
\begin{equation}\label{disk}
\rho_d (r,z)\, = \rho_0 \frac{a_d^3}{(r^2+a_d^2)^{3/2}}\, \, {\rm
sech}^2(z/z_0),
\end{equation}
where $(r,z)$ are the galactocentric cylindrical coordinates, and $a_d$ 
and $z_0$ are the disk scaling factors. The central density is 
determined by $\rho_{0} = M_d/4\pi a_d^2 z_0$ with the total mass $M_d$ 
of the disk.

We introduce model spiral galaxies of various types: late-type (L) or 
early-type (E), subdivided according to their total mass: massive (M) 
or low-mass (m). The L-types have very low bulge-to-disk mass ratios 
compared to E-types, and M-types have a massive halo component compared 
to m-types. Thus we employ four models: LM, Lm, EM, and Em. The values 
of the model parameters are given in Table~\ref{table_galaxytypes}. The 
scale height of the disk in all models is $z_0=0.25$ kpc.

\begin{table}[t]
\centering
\begin{tabular}{lclcccc}
\hline
\hline
 & & & LM & Lm & EM & Em \\
\hline \rule{0pt}{2.6ex}
disk:  & $M_{d}$ & $(10^{10}$ M$_{\odot})$ & 8.6  & 4    & 3.1    & 1.5    \\
       & $a_{d}$ & (kpc)              & 4    & 4    & 6    & 6    \\

\rule{0pt}{2.6ex}
bulge: & $M_{b}$ & $(10^{10}$ M$_{\odot})$ & 1.3  & 0.5  & 7.3    &3.5    \\
      & $a_{b}$ & (kpc)              & 0.5  & 0.1 & 1  & 1  \\

\rule{0pt}{2.6ex}
halo:  & $M_{h}$ & $(10^{10}$ M$_{\odot})$ & 42   & 14  & 49   & 16.4   \\
       & $a_{h}$ & (kpc)              & 20   & 20   & 25   & 25   \\
\hline
\end{tabular}
\caption{\small The disk, bulge, and halo parameters for late (L)
or early (E), massive (M) or low-mass (m) type models of a spiral
galaxy. }\label{table_galaxytypes}
\end{table}

\begin{figure*}[t]
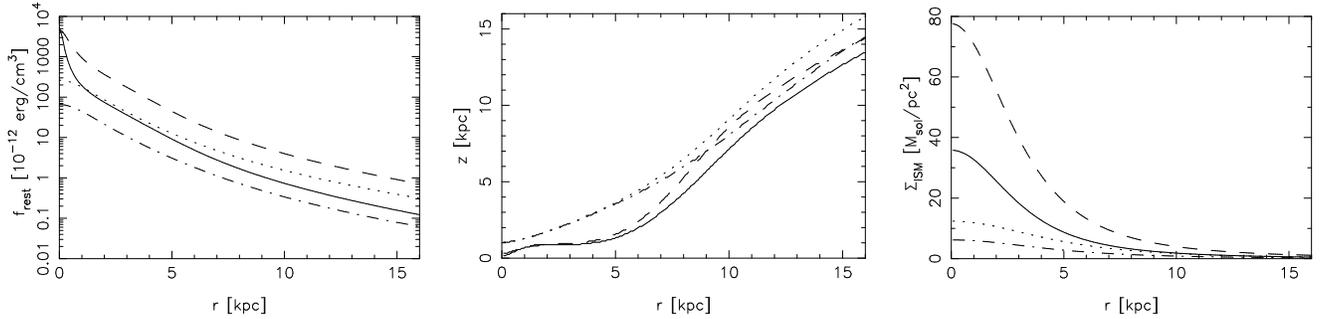

\centering
\includegraphics[width=0.3\textwidth,angle=0]{frest_r.ps}
\hspace{0.25cm}
\includegraphics[width=0.3\textwidth,angle=0]{frest_RZ_maxima.ps}
\hspace{0.25cm}
\includegraphics[width=0.3\textwidth,angle=0]{sigISM.ps}
\caption{\small Left: Maximum restoring force over $z$ as a
function of the galactocentric distance $r$ for the four galaxy
models adopted: LM (dashed), Lm (full), EM (dotted), and Em
(dash-dotted). Center: Positions of the maximum values of the
restoring force behind the disk plane for different radii $r$.
Right: Surface density $\Sigma_{\rm ISM}$ of the ISM as a function of
the galactocentric distance $r$.
}\label{frestLE}
\end{figure*}

Our standard galaxy model is the LM-type. Its rotation curve, the 
escape velocity, and the gravitational restoring force $f_{rest} =
\Sigma_{\rm ISM} \times \partial\Phi(r, z) / \partial z$ are shown in 
Fig.\ref{frest}, where $\Phi(r,z)$ is the total gravitational potential 
including all components of the galaxy, and $\partial\Phi / \partial z$ 
the total force attracting ISM elements towards the galaxy's symmetry 
plane. The escape velocity is obtained as a function $(r,z)$ from the 
potential: $v_{\rm esc}(r,z) = \sqrt{2|\Phi (r,z)|}$, and its maximum 
is always in the symmetry plane $z=0$. Rotation velocities of other 
three models, Lm, EM, and Em, are about 160 km s$^{-1}$, 220 km 
s$^{-1}$, and 150 km s$^{-1}$, respectively.

At a given $r$, the restoring force has a maximum at some distance $z$
from the symmetry plane. With increasing radius $r$, this maximum is 
farther and farther from the $z=0$ plane and its values decrease. In 
Fig.~\ref{frestLE} (left panel) we plot the maxima of the restoring 
force as a function of $r$ for different galaxy types. The LM-type 
galaxy has a massive disk and massive halo, which gives the highest 
restoring force compared to other galaxy types. Both the disk and halo 
are less massive for the Lm-type, consequently, the restoring force is 
smaller; only in the central part, where the very concentrated bulge 
dominates, the restoring force increases rapidly. The bulge is much 
more massive and extended in an EM-type galaxy, so the restoring force 
in the outer galaxy parts overcomes that of an Lm-type. For radii 
between 1 and 4 kpc, galaxy models Lm and EM show very similar runs of 
the maxima of the restoring force with $r$, but the maxima occur at 
lower $z$ in the Lm-type (see Fig.~\ref{frestLE}, central panel). The 
Em-type galaxy, where the disk and halo have lower masses compared to 
the other types, has the weakest restoring force.

We assume that the ISM initially follows the density distribution of
Eq.~\ref{disk} and that it amounts to 10\% of the total disk mass. The 
ISM surface density $\Sigma_{\rm ISM}$ is plotted as a function of the
galactocentric distance for the four galaxy types in the right hand 
panel of Fig.~\ref{frestLE}.

\section{The model of a galaxy cluster}\label{sec:cluster}
The hot ICM was observed with X-ray satellites ROSAT and ASCA and later 
resolved spatially with satellites XMM-Newton and Chandra (B\" ohringer 
2004). The large-scale ICM distribution in clusters is described with a 
$\beta$-profile (Cavaliere \& Fusco-Femiano 1976; Schindler et 
al.~1999):
\begin{equation}\label{bicm}
\rho_{\rm ICM} = \rho_{\rm 0,ICM} \left(1 + {R^2 \over R_{\rm c,ICM}^2}
\right)^{-{3 \over 2} \beta_{\rm ICM}},
\end{equation}
where $\rho_{\rm ICM}$ denotes the ICM volume density at distance $R$ 
from the cluster center, $\rho_{\rm 0,ICM}$ is the volume density of 
the ICM in the cluster center, $R_{\rm c,ICM}$ a parameter of the ICM 
central concentration, and $\beta_{\rm ICM}$ the slope parameter, which 
we set according to Schindler et al.~(1999) to 1/2. Our standard 
cluster model has $\rho_{\rm 0,ICM}=4\ 10^{-3}$ cm$^{-3}$ and $R_{\rm 
c,ICM} = 13.4$ kpc. Vollmer et al.~(2001) use a ten-times higher value 
of $\rho_{\rm 0,ICM}$ to represent the ICM distribution in the Virgo 
cluster. We discuss the difference in $\rho_{\rm 0,ICM}$ setting in 
Sect.~\ref{sec:Vollmer2001}.

\begin{figure}[t]
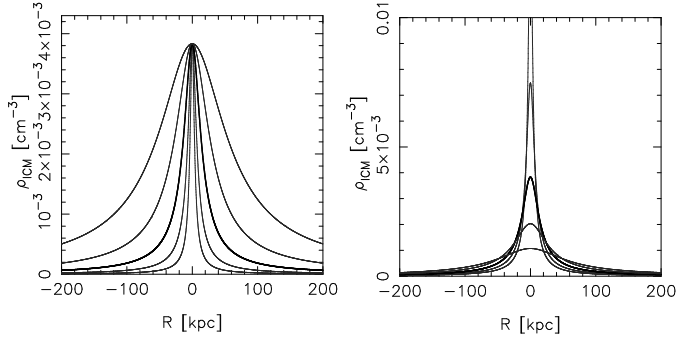

\centering
\includegraphics[height=0.24\textwidth,angle=270]{rhoICM_r_constRho.ps}
\includegraphics[height=0.24\textwidth,angle=270]{rhoICM_r_constSigma.ps}
\caption{\small
Left: Central view of the ICM density distribution in clusters with
$R_{\rm c,ICM}$ equal to 0.25, 0.5, 1, 2, 4 $\times$ 13.4 kpc, and 
$\rho_{\rm 0,ICM}=4\ 10^{-3}$ cm$^{-3}$. Right: Peaks of the ICM 
density distribution with the following combinations of the 
$\beta$-profile parameters ($R_{\rm c,ICM}$, $\rho_{\rm 0,ICM}$): 
(0.25, 3.82), (0.5, 1.95), (1, 1), (2, 0.53), and (4, 0.28) $\times$ 
(13.4 kpc, 4 10$^{-3}$cm$^{-3}$). A galaxy traversing such clusters on 
a completely radial orbit from 1 Mpc distance crosses the same amount 
of the ICM.
}\label{pram}
\end{figure}

To model either a rich galaxy cluster with a lot of hot ICM or a poor 
cluster with just a little ICM in the center, we vary the values of 
$R_{\rm c,ICM}$ and $\rho_{\rm 0,ICM}$. We multiply either standard 
value with factors of 0.25, 0.5, 1, 2, and 4. The left panel of 
Fig.~\ref{pram} illustrates the effect of varying $R_{\rm c,ICM}$ while 
$\rho_{\rm 0,ICM}$ stays fixed. The width of the ICM density peak grows 
with increasing $R_{\rm c,ICM}$.

When the galaxy flies through the cluster, it encounters a certain 
amount of ICM. The right panel of Fig.~\ref{pram} corresponds to five 
clusters with combinations of \(R_{\rm c,ICM}\), $\rho_{\rm 0,ICM}$ 
that make the galaxy cross the same amount of ICM when moving on a 
completely radial orbit, starting from a distance of 1 Mpc. The shape 
of the peak changes from narrow and high to wide and low.

The gravitational potential of a galaxy cluster is produced by the 
distributions of ICM and dark matter (DM) in the cluster. The DM is 
introduced through another $\beta$-profile. We set $R_{c,DM} = 320$ 
kpc, $\rho_{0,DM} = 1.6\ 10^{-2}$ cm$^{-3}$, and $\beta_{DM} = 1$ as 
in Vollmer et al.~(2001). The different steepness of gas and galaxy 
density profiles ($\beta_{\rm ICM}$ versus $\beta_{DM}$) corresponds 
closely to observations, with the ICM density profile steeper than the 
galaxy distribution in the inner part of the subclump. This implies DM 
of either 3.7 $10^{12}$ M$_\odot$ or 1.4 $10^{14}$ M$_\odot$ within 140 
kpc (see later) or 1 Mpc from the cluster center.  Correspondingly, 
depending on the type of the cluster, ICM represents from 0.05\% to 
30\% of the mass within the central 140 kpc (see 
Table~\ref{tab_simulations}).

\section{Initial conditions}\label{sec:initcond}
The three components of the model galaxy are represented by a number of 
particles of equal mass: 12\,000 for the halo, 6\,000 for the bulge, 
and 12\,000 for the stellar disk. Their initial distributions follow 
the density profiles of Eqs.~\ref{plummer} and \ref{disk}. The 
components are cut off at 40 kpc, 4 kpc, and 16 kpc, in the case of the 
halo, bulge, and disk, respectively. The ISM is represented with 
12\,000 SPH particles of total mass equal to 10\% of the total disk 
mass. This means that the mass of an ISM particle is in the range of 
(1.5 -- 5.7) 10$^5$ M$_\odot$ in the four introduced galaxy types. 
Their initial distribution follows the density profile of the stellar 
disk (Eq.~\ref{disk}; see right panel of Fig.~\ref{frestLE}).

The ICM of the cluster is modeled with 120\,000 SPH particles following 
the $\beta$-profile (Eq.~\ref{bicm}) with a cut-off at $R=140$ kpc from 
the center of the cluster. The total mass of the ICM particles is 
derived from cluster parameters. In the case of the standard cluster, 
there is $6\ 10^{10}$ M$_{\odot}$ of the ICM within the central 140 
kpc. Consequently the mass of an individual ICM particle is $\sim 5\ 
10^5$ M$_\odot$, which is comparable to the mass of an ISM particle. 
For the LM-type galaxy and different clusters, the mass ratio of ISM to 
ICM particles ranges between 30 and 0.04.

\begin{figure}[t]
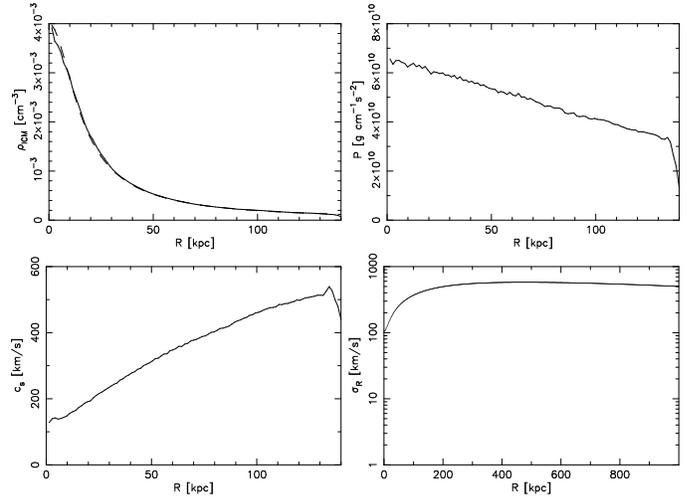

\centering
\includegraphics[width=0.24\textwidth,angle=0]{density_1.5.ps}
\includegraphics[width=0.24\textwidth,angle=0]{pressure_1.5.ps}
\includegraphics[width=0.24\textwidth,angle=0]{csnd_1.5.ps}
\includegraphics[width=0.24\textwidth,angle=0]{sigma_paper.ps}
\caption{\small 
Radial profiles of the ICM density, pressure, and speed of sound about 
0.5 Gyr before the galaxy enters the ICM distribution. The dashed curve 
in the density plot corresponds to Eq.~\ref{bicm}. Lower right: Initial 
profile of the ICM velocity dispersion $\sigma_R$, Eq.~\ref{eq_disp}.
}\label{ICMinit}
\end{figure}

The cut-off radius of the ICM particles distribution set to 140 kpc 
corresponds in the case of the standard model to the radius at which 
outer parts of the galaxy arriving from the cluster periphery start to 
be affected by steeply rising ram pressure of the crossed ICM. It means 
that we restrict the active sphere of the ICM distribution to central 
regions of the cluster, where the effect of the ICM-ISM interaction is 
strongest. At larger distances only the gravitational effect of the ICM 
to the galaxy is considered.

A more realistic simulation would spread the ICM particles throughout 
the cluster so that the galaxy runs all its orbital time through the 
medium. However, as noted by Abadi el al.~(1999), ICM particles that 
are too massive may become bullets punching holes into the ISM disk. To 
prevent this, the masses of individual ICM particles should at most be 
comparable to that of ISM particles. With this assumption and the 
number of ISM particles set to 12\,000, the sphere of 1 Mpc radius of 
the cluster would need about two million particles of the ICM. Then
the calculations would become too time-consuming and complex.

Velocity distribution of the ICM particles is calculated from second 
moment of the collisionless Boltzmann equation providing isotropy of 
the model (see e.g. Hernquist 1993):
\begin{equation}\label{eq_disp}
\sigma_R^2 = \frac{1}{\rho\,(R)} \int_R^{\infty} \rho\,(R)
\frac{d\Phi_{tot}}{dR} dR,
\end{equation}
where $\Phi_{tot}$ represents the total potential of the cluster's DM 
and ICM, and $\rho\,(R)$ is the ICM density. The $\sigma_R$ profile is 
shown in the lower right panel of Fig.~\ref{ICMinit}. 
Equation~\ref{eq_disp} is in fact an equation for the hydrostatic 
equilibrium of the cluster. After a slight initial relaxation, the 
system reaches a stable state, well before the galaxy enters the ICM 
distribution. Figure~\ref{ICMinit} depicts radial profiles of the ICM 
density (compared to the analytic profile of Eq.~\ref{bicm}), pressure, 
and speed of sound about 0.5 Gyr before the galaxy arrives. To keep ICM 
particles within the 140 kpc radius, periodic boundary conditions are 
introduced such that, when a particle reaches the edge, it continues 
its course at centrally opposite positions, with preserved velocity 
vector. Particles of the ICM are treated adiabatically, while particles 
of the ISM isothermally with $T_{\rm ISM}=10^4$ K.

\begin{figure}[t]
\centering
\includegraphics[height=0.48\textwidth,angle=270]{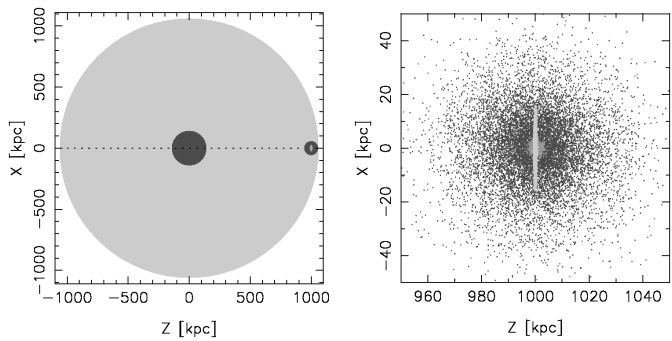}
\caption{\small
Initial configuration. Left: Galaxy falls face-on from 1 Mpc distance 
(dashed). The cut-off radius of the ICM is 140 kpc (dark), which 
outside (light) it operates only gravitationally. Right: Zoom on the 
galaxy with DM halo (dots) and disk+bulge (shaded) components.
}\label{ini}
\end{figure}

\begin{table}[t]
\centering
\begin{tabular}{cccccc}
\hline\hline \rule{0pt}{2.6ex}
$N_{\rm ICM}$ & $R_{\rm c,ICM}$ & $\rho_{\rm 0,ICM}$ & $T_{R=140\,kpc}$ & $T_0$ & $v_0$  \\
	      & (kpc)           & (cm$^{-3}$)        & Gyr              & Gyr   & (km/s) \\
\hline \rule{0pt}{2.6ex}
120\,000      & 13.4            & 4 10$^{-3}$        & 1.53             & 1.64  & 1260   \\
\hline
\end{tabular}
\caption{\small
Parameters of the standard simulation run. Times at which galaxy enters
the ICM and reaches the cluster center are stated, together with its 
velocity in the cluster center.}
\label{tab_std}
\end{table}

Figure~\ref{ini} shows the initial configuration of the system. Galaxy 
in face-on orientation relative to the orbit is located at the cluster 
periphery at 1 Mpc distance from the center, {\boldmath$R$} $= (x,y,z) 
= [0,0,1]$ Mpc, with no systematic velocity. Consequently, it falls 
freely towards the cluster center. Thus, in this work, all the orbits 
are completely radial, free-fall orbits from the distance $R$ = 1 Mpc 
to the cluster center. For the standard cluster, the galaxy reaches the 
cluster center after about 1.64 Gyr, at a velocity of about 1300 km 
s$^{-1}$. It means that it moves very supersonically through the ICM in 
the central regions (cf. Fig.~\ref{ICMinit}), and we expect a strong 
bow shock to form in front of the galactic disk. Domainko et al.~(2006) 
draw attention to the role of the bow shock: the ram pressure behind 
the shock is lower since the relative velocities of the ICM particles 
are smaller there than in front of the shock. On the contrary, 
Rasmussen et al.~(2006) notice that, while the ram pressure is reduced 
at the shock, \(P_{th}+\rho v^2\), where $P_{th}$ is the thermal ICM 
pressure behind the shock, stays conserved for an inviscid fluid, and 
thus the reduction of the ram pressure behind the shock is balanced by 
an increase in static thermal pressure. Therefore, the force per unit 
area acting on the disk is close to the pre-shock ram pressure.

Table~\ref{tab_std} summarizes the parameters of the standard 
simulation run. Along the radial orbit, we can compute the ram pressure 
$p_{ram}(R) = \rho_{\rm ICM} v_{gal}^2$ acting on the galaxy as it 
crosses the ICM. Ram pressure always peaks in the cluster center where 
the ICM density is highest, and also where the orbital velocity has its 
maximum. Following Fig.~\ref{pram}, for broader clusters the increase 
in the ram pressure starts at larger distances from the cluster center. 
With the same central density $\rho_{\rm 0,ICM}$ the peak value of 
$p_{ram}$ is slightly higher for broader clusters due to the higher 
contribution of the ICM to the overall cluster gravity 
(Fig.~\ref{pram_rc_rho}, left panel).

In the simulations, we follow the galaxy on its orbit during 2 Gyr. 
From the very beginning at 1 Mpc distance, the galaxy itself evolves 
and thus instabilities and spiral arms form. Consequently, the galaxy 
enters the ICM sphere at about 1.5 Gyr as dynamically evolved.

\section{Simulation method}\label{sec:method}
We performed 3D tree/SPH N-body simulations using GADGET in version
1.1 (Springel et al. 2001).  The tree method is a highly effective 
Lagrangian technique for computing gravity forces in non-homogeneous 
systems. The particles are arranged into groups hierarchically ordered 
from the smallest ones containing only one particle up to the biggest
one embracing the whole system. The scheme of an oct-tree structure 
dividing the computational space into a sequence of cubes is used 
(Barnes \& Hut 1986). Then, the gravitational force exerted by distant 
groups on a particle is approximated by their lowest multipole moments, 
which reduces the computational costs to order $O(N \log N)$.

The smoothed particle hydrodynamics (SPH, Lucy 1977, Gingold \& 
Monaghan 1977) is a Lagrangian method for solving hydrodynamical 
problems in which real fluids are represented by a set of fluid
elements, i.e. particles. The key idea of the SPH is that every 
particle has an effective radius within which hydrodynamical 
interactions with neighboring particles operate. Thus the particles are 
no longer point masses but are smoothed out, and local hydrodynamical 
properties are interpolated from the local mass distribution. The 
smoothing sizes of particles are determined to correspond to a radius 
within which a given number of neighboring particles is contained.

GADGET in version 1.1 resolves only one gaseous phase, and all gas 
particles occurring in the system are hydrodynamically treated as one 
group. However, the volume densities of the ISM in the disk and of the 
surrounding ICM differ by several orders of magnitude, as well as their
temperatures. To treat their interaction correctly, the number 
densities of the ICM and ISM particles, $n_{\rm ICM}$ and $n_{\rm 
ISM}$, should be comparable. Otherwise, the ICM particles located far 
from the galaxy would be larger in the SPH sense than the ISM 
particles, since the number of neighboring particles is kept constant 
throughout the system. When these large ICM particles would later 
approach the disk, their sizes would be re-evaluated from the local 
mixing of ICM and ISM particles, and artificial punching effects could 
follow. 

\begin{figure}[t]
\centering
\includegraphics[height=0.48\textwidth,angle=270]{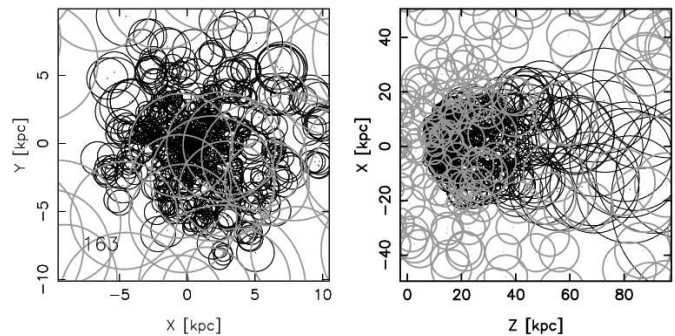}
\caption{\small
SPH sizes of the ISM (black) and ICM (grey) particles at the epoch when 
the galaxy runs through the cluster center. $N_{\rm ICM}=120\,000$, 
$N_{\rm ISM}=12\,000$. Left: Face-on zoom view of the galaxy 
surroundings with the ISM particles located within $\pm$ 2 kpc about 
the disk plane, and the ICM particles within $\pm$ 20 kpc about the 
disk plane. Right: Edge-on view of the situation with ICM particles 
located within $\pm$ 20 kpc about the disk in the line of sight 
direction. Only every tenth particle is displayed.
}\label{smooth}
\end{figure}

\begin{figure*}[t]
\centering
\includegraphics[width=0.98\textwidth,angle=0]{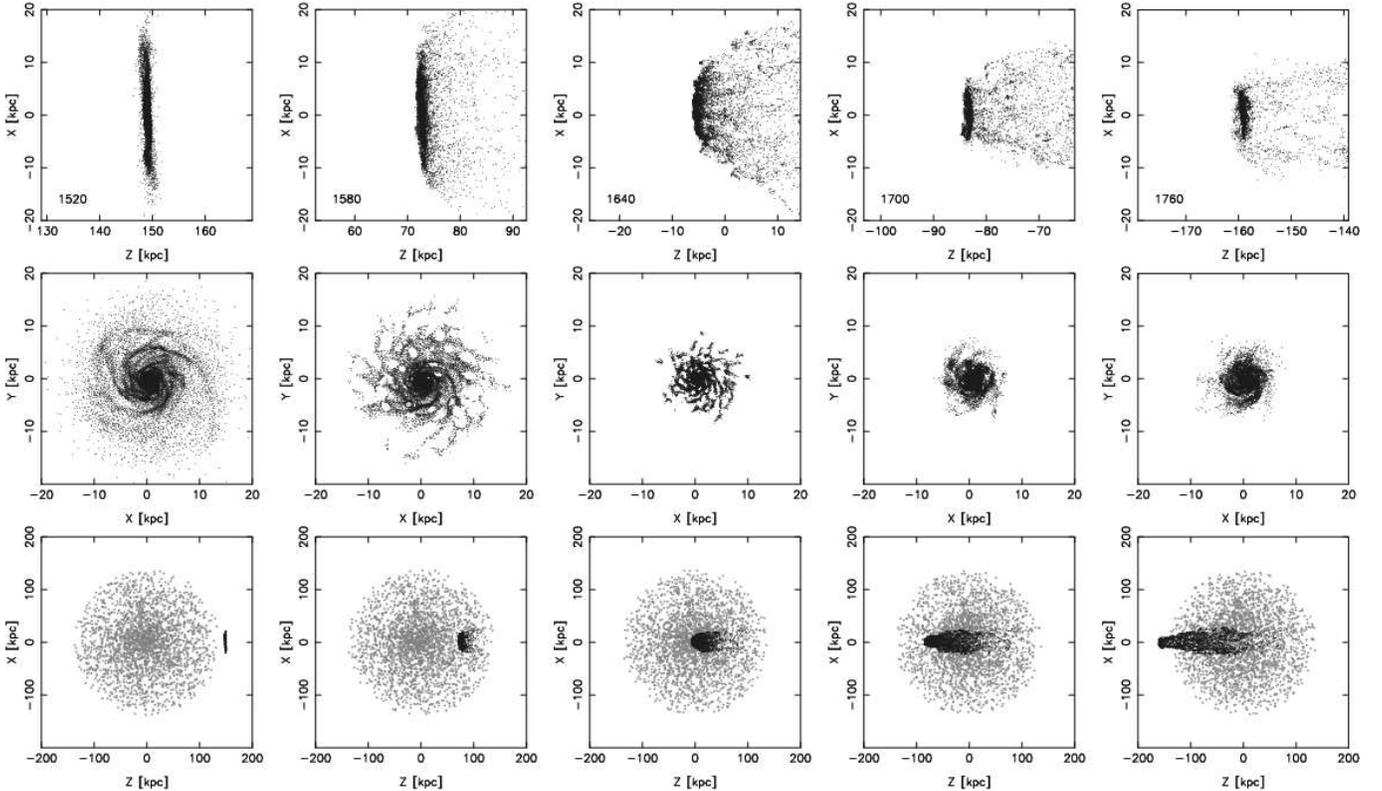}
\caption{\small
Time sequence -- a galaxy crossing the cluster center. From top down: 
Edge-on projection and face-on cut ($|z|<\pm$ 2 kpc) of the ISM particles; 
the tail of the stripped ISM in the central distribution of the ICM 
particles. The time since the beginning of the free fall of the galaxy 
from the cluster edge is given in Myr in the first row.
}\label{LM_simul}
\end{figure*}

To achieve similar number densities of both gases, a huge number of ICM 
particles are necessary. With these constraints in mind, we introduce a 
set of changes into GADGET related to the fact that we need to simulate 
two different gaseous phases (ICM and ISM) with different spatial 
resolution. First, the subroutine for searching the neighboring 
particles is adapted to distinguish between the two phases: the 
smoothing radii $h$'s of either ICM or ISM particles are calculated 
separately from the neighbors of a corresponding phase. In other words, 
the search continues until $N_b$ proper neighbors are found, while 
those of the second phase are omitted. Then a reasonable number of ICM 
particles, that are consequently substantially larger than the ISM, can 
be used, and their size does not change even when interacting with the 
disk. It ensures that all the disk area is covered by the ICM 
particles. Second, the hydrodynamical interaction of the two phases can 
proceed: Both phases are searched together to find all particles with 
$|${\boldmath$r$}$_{ij}| < \max(h_i,h_j)$. A given ISM particle then 
interacts with all the ICM particles within which smoothing radii it is 
contained. Third, the adiabatical and isothermal treatments of the ICM 
and ISM are implemented, respectively.

Figure~\ref{smooth} compares the ISM (black circles) versus ICM (grey
circles) particle sizes in the surroundings of the galaxy running 
through the center of the standard cluster. Face-on and edge-on 
snapshots are displayed, and the $N_{\rm ICM}=120\,000$ and $N_{\rm 
ISM}=12\,000$. Only every tenth particle is plotted, which gives the 
impression of voids occurring among particles. One notes that the size 
of the ICM particles is kept constant in the disk vicinity.

To achieve a sufficient spatial resolution in the disk, we fix the 
number of ISM particles $N_{\rm ISM}$ to an adequate value. In all 
simulations, the resolution in the disk is then the same. From 
Fig.~\ref{smooth} it follows that the ICM particles are substantially
larger than their ISM counterparts, each of them covering a large part 
of the disk. That means that a group of close ISM particles lying 
within the smoothing radius of a given ICM particle feel only the ram
pressure of the ICM but no pressure gradients, as would be the case
if the sizes of ICM and ISM particles were comparable. In 
section~\ref{sec:tests} (Fig.~\ref{pocty_NICM}), we discuss test runs 
with varying numbers of ICM particles.

Our simulation approach using only a reasonable number of ICM 
particles, and thus neglecting the detailed hydrodynamics in the 
ISM--ICM interaction treatment, has some similarities with the method 
of Vollmer et al.~(2001), who in their sticky-particle simulations 
include the effect of the ram pressure only analytically as an 
additional acceleration on the ISM clouds located at the surface of the 
gas distribution. In our case, of course, the interaction between ICM 
and ISM is not restricted to the surface layer.

Our approach is more realistic than Vollmer et al.~(2001) as far as the 
ICM is concerned. The ICM phase is treated fully hydrodynamically. 
Therefore, as discussed later, simulations develop a bow shock that 
forms in the ICM in front of the galaxy. Consequently, the incoming ICM 
particles are deflected from their direction to flow around the disk. 
Moreover, when (from the SPH point of view) small ISM particles are 
stripped out of the disk, their number density falls and their size 
increases. In the tail of stripped particles, the sizes of both the ISM 
and ICM then become comparable, and the hydrodynamical treatment of the 
tail is correct (see Fig.~\ref{smooth}).

\section{Simulation results}\label{sec:results}
First we describe results of the standard simulation run (see 
Table~\ref{tab_std}) of the LM galaxy. This shows how the ISM of the
galaxy changes under the influence of a) the ram pressure due to 
relative motion of the ISM and ICM, b) the gravitational field of the 
cluster that changes along the orbit, and c) the evolving galactic 
gravitational field that includes the self-gravity of the ISM disk. We 
note that no effects of the ram pressure to the stellar disk, bulge or 
halo galaxy components are observed in the simulations described below. 

\subsection{LM galaxy in the standard cluster}\label{LMinstand}
Figure~\ref{LM_simul} displays snapshots of the ISM disk at five epochs
of the galaxy's crossing through the central parts of the cluster. 
Freely falling from the 1 Mpc distance, it approaches, at time 1.52, 
the edge of the ICM particle distribution, where ram pressure rises. At 
1.64 Gyr it passes through the cluster center and experiences the peak 
ram pressure. At 1.76 Gyr it leaves the ICM on the other side of its 
distribution, relaxes after the fade-out of the ram pressure, and 
further continues towards the second apocenter. Figure~\ref{LM_simul} 
shows edge-on and face-on views of the ISM disk, and a wide view of the 
central parts of the cluster with the ICM particles.

\begin{figure}[t]
\centering
\includegraphics[height=0.4\textwidth,angle=270]{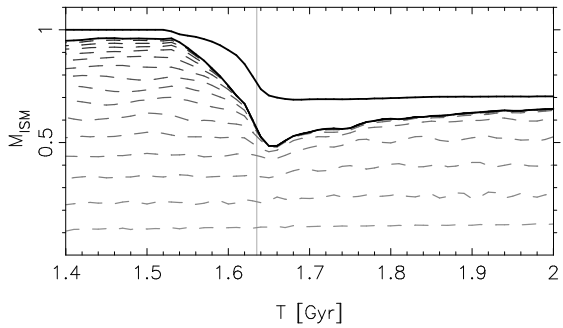}
\caption{\small
The standard simulation run. Time evolution of the ISM mass within 16
different disk radii (1 -- 16 kpc) and $|z|<1$ kpc (dashed curves). The
uppermost curve corresponds to the ISM mass bound to the galaxy. The
instant of crossing the cluster center is marked with the vertical line.
}\label{std}
\end{figure}

When the galaxy enters the ICM distribution, the local ram pressure 
affects only the outer disk regions. The ISM from these regions is 
stripped and thus forms the end of the tail of stripped gas. Later on, 
more inner parts also get stripped by the increasing ram pressure, 
filling the central portion of the tail that is clearly visible in 
Fig.~\ref{LM_simul}. The opening angle of the tail changes with time, 
which reflects the time evolution of the ram pressure vs. restoring 
force contest. Schulz \& Struck (2001) describe one effect of the ram 
pressure enhancing the spiral structure of the disk since particles 
from inter-arm positions with lower surface densities are more easily 
removed. This effect is observed at time 1.58 Gyr.

The process of stripping is displayed in Fig.~\ref{std}. It shows the
number of ISM particles, i.e. the ISM mass, within different disk radii
growing from 1 kpc to 16 kpc with a step of 1 kpc, and within a $|z|<1$ 
kpc belt about the disk plane as a function of the orbital time. The
uppermost curve shows the ISM mass $M_{\rm bnd}$ that is bound to the 
galaxy, i.e. the number of ISM particles with negative total energy. 
Before the galaxy enters the ICM zone at $\sim 1.53$ Gyr, no important 
changes are observed. Once the ram pressure rises, particles from 
decreasing radii are shifted out of the evaluation belt. With a delay 
of about 20 Myr after the galaxy's passage through the center, a 
maximum amount of the ISM is shifted out of the $|z|<1$ kpc zone, and 
curves in Fig.~\ref{std} show a minimum corresponding to 49\% of the
initial total ISM mass. Particles down to about 4 kpc are influenced.
After the minimum, a strong reaccretion of the ISM previously only 
shifted out of the disk by a short-term effect of the strongest ram 
pressure, but staying bound to the galaxy occurs. Of course, both the 
minimum value and the amount of the reaccreted material depend on the 
definition of the evaluation zone.

With a constant flow of the ICM, Roediger \& Hensler (2005) identify 
three phases of stripping: (i) an instantaneous phase when the outer 
disk parts are bent in the wind direction but stay bound to the galaxy; 
(ii) a dynamic intermediate phase, during which the bending breaks, a 
part of the ISM is stripped, and another part falls back to the disk; 
and (iii) a quasi-stable continuous viscous stripping phase when the 
outer disk layers are peeled off by the Kelvin-Helmholz instability at 
a rate of $\sim$ 1 M$_\odot$ yr$^{-1}$. The duration of the first phase 
is $\sim$ 20 -- 200 Myr, while the following second phase is about ten 
times longer. In our simulations, we observe behavior similar to the 
first two phases; however, the evolution of the $M_{\rm bnd}(t)$ and 
$M_{|z|<1\rm kpc}(t)$ under the operation of a peaked ram pressure is 
different (see Fig.~\ref{std}). Once the ram pressure rises, the outer 
disk layers are released, while the inner ones are only shifted from 
the $|z|<1$ kpc zone. Although the rate of the release is lower 
compared to the shifting, both effects occur simultaneously. When the 
ram pressure ceases, the shifted, but bound, ISM reaccretes. 

From Fig.~\ref{std} we determine the final post-stripping mass $M_{\rm 
final}$ of the galaxy as the mass of bound ISM at the final simulation 
time of 2 Gyr: $M_{\rm final}=0.71 M_{\rm d,ISM}$. The stripping radius 
$r_{strip}$ is trickier to measure, since all the reaccreting material 
is not back in the disk at the final time simulated. We estimate it to 
be $r_{strip}\sim 6$ kpc. Initially in Fig.~\ref{std} about 5\% of the 
12 000 ISM particles are, due to relaxation processes and scattering on 
spiral arms, already outside the region $r<16$ kpc, $|z|<1$ kpc before 
the ram pressure starts to play a role. But all these particles have 
negative total energy. 

\begin{figure*}[t]
\centering
\includegraphics[height=0.98\textwidth,angle=270]{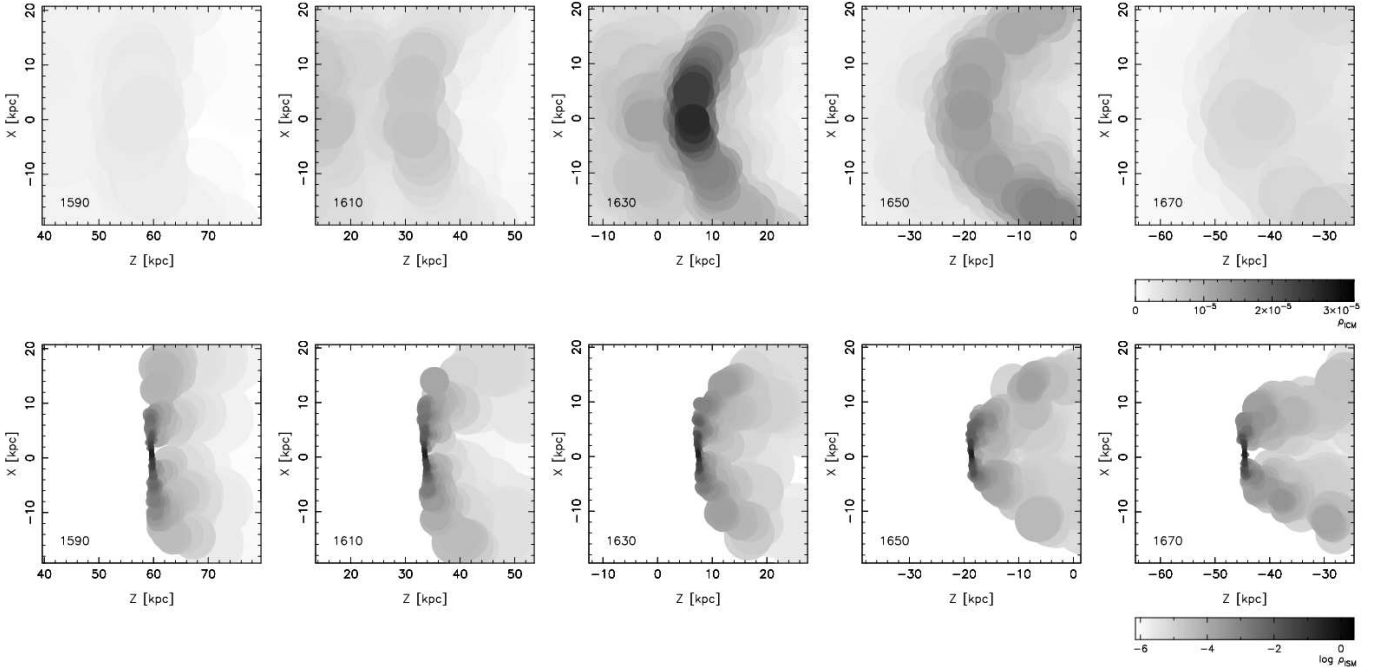}
\caption{\small
Density of the ICM in the surroundings of the LM-type galaxy (top) and 
of the ISM (bottom). A bow shock forms in front of the disk. Individual
particles are displayed as filled circles with radii equal to their SPH
smoothing sizes $h$'s. The scale of the ISM density is logarithmic due
to large differences in the disk plane and in the tail of stripped 
particles. The time in Myr is given in the left lower corner of frames.
}\label{den}
\end{figure*}

\begin{figure*}[t]
\centering
\includegraphics[height=0.98\textwidth,angle=270]{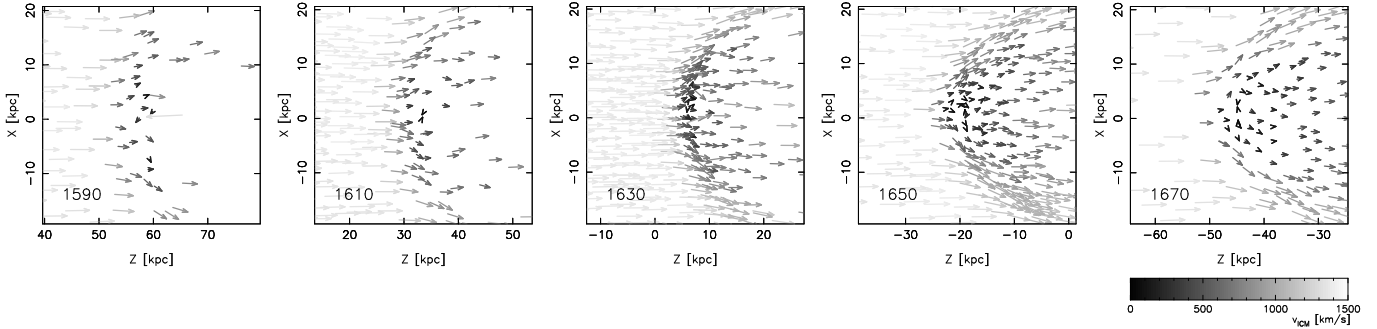}
\caption{\small
The flow of ICM particles around the LM-type galaxy. Vectors of the 
relative velocity of ICM particles interacting with the ISM disk are
displayed.
}\label{velo}
\end{figure*}

The upper row of Fig.~\ref{den} shows the ICM density distribution in
the surroundings of the galaxy passing the densest parts of the cluster 
at times 1.59 -- 1.67 Gyr. The particles are displayed as filled 
circles with radii equal to their SPH smoothing lengths $h$'s, where 
the gray shade corresponds to the density. Due to the supersonic 
motion, we see a bow shock forming in the ICM. It is most pronounced 
when the galaxy passes through the very center of the cluster. The 
bottom row of Fig.~\ref{den} shows the ISM density. Due to large 
differences in the density between the disk and the tail, the 
grey-scale is logarithmic. Note the growth of ISM particles when they 
are stripped from the disk. In the Appendix, Fig.~\ref{den480} shows 
the ICM and ISM density distributions for a simulation with $N_{\rm 
ICM}=480\,000$.

Figure~\ref{velo} shows the velocity vectors of the ICM particles 
interacting with the ISM disk. At the surface of the disk, ICM 
decelerates from more than 1000 km s$^{-1}$ to less than 100 km 
s$^{-1}$. We can also see the deflection of the ICM flow to the sides 
along the shock, and a low-velocity ICM in the shadow behind the 
galaxy. In Fig.~\ref{velo480} of the Appendix the ICM velocity field 
for the case of $N_{\rm ICM}=480\,000$ is shown.

\subsection{LM galaxy in various clusters}
As we have seen in Fig.~\ref{std}, the stripping of the ISM is a 
dynamical process depending strongly on the shape of the ram pressure
peak. Therefore, we present the results of a set of twenty-five
simulations of the LM-type galaxy crossing clusters with varying
distributions of the ICM. 

\subsubsection{Results from the simulations}
\begin{figure}[t]
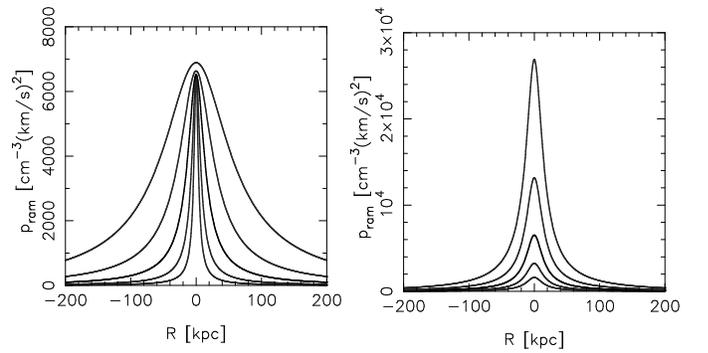

\centering
\includegraphics[height=0.24\textwidth,angle=270]{pram_r_constRho.ps}
\includegraphics[height=0.24\textwidth,angle=270]{pram_r_constRc.ps}
\caption{\small
Central views of the ram pressure peaks with a fixed $\rho_{\rm 
0,ICM}=4\ 10^{-3}$ cm$^{-3}$ and $R_{\rm c,ICM}=$ 0.25, 0.5, 1, 2, 4 
$\times$ 13.4 kpc (left), and a fixed $R_{\rm c,ICM}=$ 13.4 kpc and 
$\rho_{\rm 0,ICM}=$ 0.25, 0.5, 1, 2, 4 $\times$ $4\ 10^{-3}$ cm$^{-3}$ 
(right).
}\label{pram_rc_rho}
\end{figure}

\begin{figure*}[t]
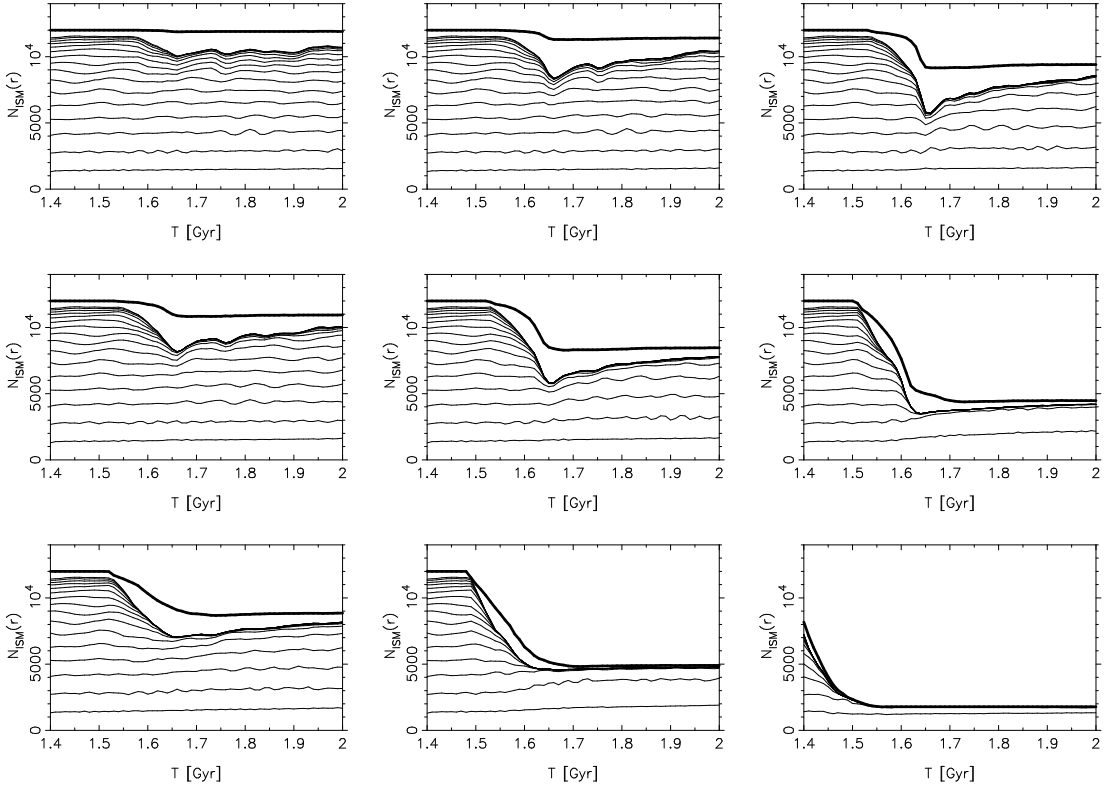

\centering
\includegraphics[height=0.25\textwidth,angle=270]{egyRc0.25Rho0.25.ps}
\hspace{0.25cm}
\includegraphics[height=0.25\textwidth,angle=270]{egyRc0.25Rho1.ps}
\hspace{0.25cm}
\includegraphics[height=0.25\textwidth,angle=270]{egyRc0.25Rho4.ps}\\~\\
\includegraphics[height=0.25\textwidth,angle=270]{egyRc1Rho0.25.ps}
\hspace{0.25cm}
\includegraphics[height=0.25\textwidth,angle=270]{egyRc1Rho1.ps}
\hspace{0.25cm}
\includegraphics[height=0.25\textwidth,angle=270]{egyRc1Rho4.ps}\\~\\
\includegraphics[height=0.25\textwidth,angle=270]{egyRc4Rho0.25.ps}
\hspace{0.25cm}
\includegraphics[height=0.25\textwidth,angle=270]{egyRc4Rho1.ps}
\hspace{0.25cm}
\includegraphics[height=0.25\textwidth,angle=270]{egyRc4Rho4.ps}
\caption{\small
Stripping of the LM-type galaxy crossing clusters with varying ICM 
distribution: from left to right: $\rho_{\rm 0,ICM}=$ 0.25, 1, 4 
$\times$ 4 10$^{-3}$ cm$^{-3}$, from top to bottom: $R_{\rm c,ICM}=$ 
0.25, 1, 4 $\times$ 13.4 kpc. See Fig.~\ref{std} for details. 
Simulations in the same row thus have a constant value of the $R_{\rm 
c,ICM}$ parameter, and simulations in the same column keep the 
$\rho_{\rm 0,ICM}$ value. The central panel corresponds to 
Fig.~\ref{std}.
}\label{LM_res}
\end{figure*}

Multiplying the standard values of the parameters $R_{\rm c,ICM}$ and 
$\rho_{\rm 0,ICM}$ by factors of 0.25, 0.5, 1, 2, and 4, we study the 
effects of a varying width and height of the ICM density peak on the 
stripping results. Fig.~\ref{pram_rc_rho} illustrates the ram pressure 
profiles corresponding to the ICM distribution with varying widths but 
a constant height (left), and varying heights but a constant width 
(right). Fig.~\ref{LM_res} shows the stripping results of nine runs 
from the above set of twenty-five simulations. Individual panels are 
like those in Fig.~\ref{std} and correspond to $\rho_{\rm 0,ICM}=$ 
0.25, 1, 4 $\times$ 4 10$^{-3}$ cm$^{-3}$ (from left to right) and 
$R_{\rm c,ICM}=$ 0.25, 1, 4 $\times$ 13.4 kpc (from top to bottom). 

The main stripping event is rather short in all the cases and takes 100 
to 200 Myr. Going from the upper left corner to lower right of
Fig.~\ref{LM_res}, the stripping expands to more and more inner parts
of the galaxy. It means that both increasing width and height of the
ICM density peak support the stripping. Nevertheless, the process of
stripping itself is different in either direction:

\noindent
a) following a row, i.e. for a fixed value of the $R_{\rm c,ICM}$, the 
stripping of the ISM deepens with growing value of the central density. 
In the upper row, the ram pressure peak is narrow, with $R_{\rm c,ICM} 
= 3.4$ kpc. If the peak is not high (left), most of the ISM remains 
inside the $|z|<1$ kpc layer. About $\sim 12\%$ of the ISM has been 
shifted out of the layer, but almost all of it remains gravitationally 
bound to the galaxy and later will be reaccreted. The wiggles seen on 
the curves are $z$-oscillations of the reaccreting material around the 
disk plane. If the ram pressure pulse is high enough (right), it kicks 
the disk abruptly, which accelerates the outer ISM directly to 
$v_{esc}$ and only shifts the inner ISM out of the disk plane. 
Consequently, $\sim 52$\% of the ISM moves out of the $|z|<1$ kpc 
layer, but more than a half of it, i.e. $\sim 30$\% of the original 
amount of the ISM, remains gravitationally bound and is later 
reaccreted. It may be seen as a deep notch with the deepest part at T 
$\sim$ 1.65 Gyr.

\noindent
b) Following a column, i.e. for a fixed value of the central density,
the stripping begins in wider clusters at earlier times, since the
value of the ram pressure needed to affect the outer ISM shifts to 
greater distances from the cluster center. Thus, the minimum of the ISM
mass in the evaluation zone $|z|<1$ kpc broadens. Broad clusters, with
$R_{\rm c,ICM}=53.6$ kpc, are not only much more effective in removing 
the ISM particles outside the $|z|<1$ kpc layer, but the more extended 
ram pressure pulse is also able to accelerate more ISM and remove it 
from the parent galaxy.

Table~\ref{tab_simulations} quantifies the stripping results of the set 
of twenty-five simulations and adds values of parameters characterizing 
individual runs. One sees that in a small (low and narrow) cluster, 
poor in ICM, where $\rho_{\rm 0,ICM}=1\ 10^{-3}$ cm$^{-3}$ and $R_{\rm 
c,ICM}=3.4$ kpc, the LM-type galaxy is stripped of at most 1\% of its 
ISM. It happens between the orbital time 1.53 and 1.64 Gyr. On the other 
hand, in a large (high and wide) cluster with a lot of ICM, where 
$\rho_{\rm 0,ICM}=1.6\ 10^{-2}$ cm$^{-3}$ and $R_{\rm c,ICM} = 53.6$ 
kpc, the stripping of 85\% of the ISM is achieved between orbital times 
1.37 and 1.56 Gyr.

\begin{figure*}[t]
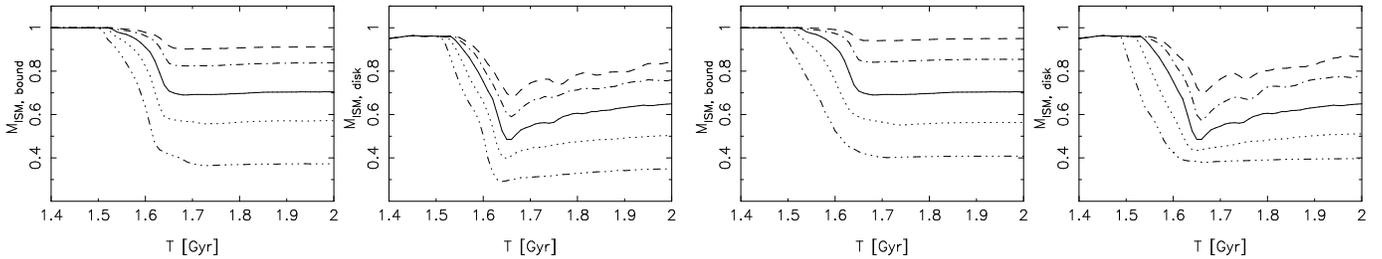

\centering
\includegraphics[width=0.24\textwidth,angle=0]{pocty_rc1_bound_BW.ps}
\includegraphics[width=0.24\textwidth,angle=0]{pocty_rc1_disk_BW.ps}
\hspace{0.1cm}
\includegraphics[width=0.24\textwidth,angle=0]{pocty_rho1_bound_BW.ps}
\includegraphics[width=0.24\textwidth,angle=0]{pocty_rho1_disk_BW.ps}
\caption{\small
Stripping of the LM-type galaxy in central parts of clusters with: 
(left pair) fixed $R_{\rm c,ICM}=13.4$ kpc and varying $\rho_{\rm 
0,ICM}$: $0.25$, $0.5$, $1$, $2$, and $4$ x $4\ 10^{-3}$ cm$^{-3}$ 
(from top to bottom) and (right pair) with fixed $\rho_{\rm 0,ICM}=4\ 
10^{-3}$ cm$^{-3}$ and varying $R_{\rm c,ICM}$: $0.25$, $0.5$, $1$, 
$2$, and $4$ x $13.4$ kpc (from top to bottom). Evolution of the ISM 
mass bound to the galaxy or staying within $r<16$ kpc and $|z|<1$ kpc 
about the disk plane is displayed. Reaccretion of the stripped ISM is 
observable.
}\label{pocty_bound_disk}
\end{figure*}

Figure~\ref{pocty_bound_disk} summarizes the results of the set of 
simulations. The general trends of galaxies crossing wider and higher 
ICM density peaks losing a growing amount of the ISM, as well as 
the reaccretion weakening, are clearly visible. In clusters with a 
constant width, the stripping starts within a short time range, while 
it occurs earlier in wide clusters, where galaxies fall faster towards 
the cluster center and the ICM has a greater extent.

\begin{table*}[t]
\centering
\begin{tabular}{ccccccccc}
\hline \hline \rule{0pt}{2.6ex}
$R_{\rm c,ICM}$ & $\rho_{\rm 0,ICM}$ & M$_{\rm ICM}^{R<140\,\rm kpc}$ & 
$\Sigma_{\rm ICM}^{R<140\,\rm kpc}$ &
$p_{ram}^{max}$ & $r_{strip}$ & $M_{strip}$ & $M_{min}$ & $M_{accr}$\\
\rule{0pt}{2.6ex}
(kpc) & (10$^{-3}$cm$^{-3}$) & (10$^{11}$M$_{\odot}$) & (M$_{\odot}$/pc$^{2}$)
& (cm$^{-3}$km$^2$s$^{-2}$) & (kpc) & (\%) & (\%) & (\%)\\
\hline \rule{0pt}{2.6ex}
     & 1  & 0.02  & 0.38  & 1\,612  & 13.8 & 1  & 84 & 15 \\
     & 2  & 0.04  & 0.76  & 3\,231  & 12.8 & 2  & 78 & 20 \\
3.4  & 4  & 0.08  & 1.53  & 6\,453  & 11.8 & 5  & 70 & 25 \\
     & 8  & 0.16  & 3.05  & 12\,949 & 10.3 & 12 & 59 & 29 \\
     & 16 & 0.32  & {\bf6.10}  & 25\,900 & 7.5  & 22 & 48 & 30 \\
\hline \rule{0pt}{2.6ex}
     & 1  & 0.06  & 0.74  & 1\,615  & 12.3 & 3  & 75 & 22 \\
     & 2  & 0.11  & 1.47  & 3\,233  & 11.2 & 7  & 68 & 25 \\
6.7  & 4  & 0.22  & 2.94  & 6\,480  & 8.7  & 14 & 57 & 29 \\
     & 8  & 0.44  & {\bf5.87}  & 13\,014 & 7.1  & 26 & 47 & 27 \\
     & 16 & 0.89  & 11.76 & 26\,229 & 4.5  & 41 & 36 & 23 \\
\hline \rule{0pt}{2.6ex}
     & 1  & 0.15  & 1.39  & 1\,618  & 10.8 & 9  & 68 & 23 \\
     & 2  & 0.30  & 2.78  & 3\,245  & 8.4  & 16 & 59 & 25 \\
13.4 & 4  & 0.61  & {\bf5.56}  & 6\,524  & 6.3  & 29 & 49 & 22 \\
     & 8  & 1.21  & 11.13 & 13\,182 & 4.1  & 43 & 40 & 17 \\
     & 16 & 2.43  & 22.26 & 26\,894 & 2.6  & 63 & 29 & 8  \\
\hline \rule{0pt}{2.6ex}
     & 1  & 0.39  & 2.56  & 1\,625  & 8.8  & 16 & 64 & 20 \\
     & 2  & 0.78  & {\bf5.12}  & 3\,272  & 7.0  & 27 & 55 & 18 \\
26.8 & 4  & 1.57  & 10.25 & 6\,631  & 5.0  & 44 & 44 & 12 \\
     & 8  & 3.14  & 20.50 & 13\,608 & 2.8  & 62 & 31 & 7  \\
     & 16 & 6.28  & 40.99 & 28\,604 & 2.5  & 80 & 20 & 0  \\
\hline \rule{0pt}{2.6ex}
     & 1  & 0.90  & {\bf4.52}  & 1\,641  & 7.7  & 26 & 59 & 15 \\
     & 2  & 1.80  & 9.04  & 3\,336  & 5.0  & 42 & 47 & 11 \\
53.6 & 4  & 3.61  & 18.08 & 6\,888  & 3.5  & 59 & 38 & 3  \\
     & 8  & 7.21  & 36.16 & 14\,637 & 2.9  & 76 & 24 & 0  \\
     & 16 & 14.42 & 72.29 & 32\,722 & 2.4  & 85 & 15 & 0  \\
\hline
\end{tabular}
\caption{\small
Summary of stripping results of the set of twenty-five simulations with
the LM-type galaxy crossing clusters with varying values of the $R_{\rm 
c,ICM}$ and $\rho_{\rm 0,ICM}$ parameters. Stripping radius $r_{strip}$ 
of the ISM disk, mass of the stripped ISM $M_{strip}$, minimum mass of
the ISM within the evaluation zone, and mass of the reaccreted ISM 
$M_{accr}$ are stated. $M_{strip}+M_{min}+M_{accr}=100\%$. Values of
peak ram pressure $p_{ram}^{max}$, column density of the encountered
ICM $\Sigma_{\rm ICM}$, and mass of the ICM within central part of 
clusters $M_{\rm ICM}$ are added.
}\label{tab_simulations}
\end{table*}

\subsubsection{Comparison with the criterion of Gunn \& Gott}
Focusing in Table~\ref{tab_simulations} on values of the column density
of the encountered ICM, we note that the passages through clusters with 
a similar amount of the encountered ICM show a similar amount of the 
stripped ISM. Figure~\ref{pocty_constSigICM} compares the process of 
stripping in five clusters with $\Sigma_{\rm ICM}\sim$ (5 -- 6) 
M$_\odot$/pc$^2$ which are in Table~\ref{tab_simulations} marked in 
boldface. In their case, $M_{strip}=$ 22\%, 26\%, 29\%, 27\%, and 26\%. 
However, the stripping history differs: in a narrow cluster with a high 
ICM density peak, there is a lot of ISM shifted out of the $|z|<1$ kpc 
zone, but this is reaccreted later. In a broad cluster with a rather 
shallow density peak, more ISM stays near the galaxy symmetry plane and 
one sees less of the reaccretion.

The two extreme cases substantially differ in the peak ram pressure; in
the first case, it is almost 16 $\times$ higher compared to the second
case. Gunn \& Gott (1972) criterion (Eq.~\ref{striprad}) would predict 
a completely different amount of stripping in these cases. However, as 
we argue later, the final stripping depends on the column density of 
the encountered ICM, which is similar in the two cases. In 
Fig.~\ref{eom_simul} a comparison of the simulation results with Gunn 
\& Gott (1972) predictions can be seen.

In the left panel of Fig.~\ref{mfinsigICM}, the dependence of the 
remaining gas mass fraction $M_{\rm final} = 1 - M_{strip}$ on the 
extent of the ICM distribution in the cluster is plotted. For 
increasing $R_{\rm c,ICM}$, the degree of stripping becomes less and 
less dependent on the size of the ICM distribution (characterized by 
the $R_{\rm c,ICM}$ itself) and approaches a constant, which depends on 
the value of $\rho_{\rm 0,ICM}$. This depends on whether or not the ISM 
is accelerated during the ram pressure pulse to the escape velocity. 
The critical size of the active region $\Delta R_c$ is $\Delta R_c = 
v_{esc}\ v\ \Sigma_{\rm ISM}/p_{\rm ram}^{\rm eff}$, where $p_{\rm 
ram}^{\rm eff}$ is an effective value of the ram pressure. Here we take 
$p_{\rm ram}^{\rm eff} = p_{\rm ram}^{\rm max}/2$. With $\Sigma_{\rm 
ISM}=10^{21}$ cm$^{-2}, v_{\rm esc}=400$ km s$^{-1}$, $p_{ram}^{\rm 
max}=6000$ cm$^{-3}$ km$^2$ s$^{-2}$, and $v=1300$ km s$^{-1}$, we get 
$\Delta R_c \approx 60$ kpc. This corresponds nicely to $R_{\rm c,ISM} 
= \Delta R_c / 2 = 30$ kpc (see Fig. \ref{mfinsigICM}), where $M_{\rm 
final}$ loses its dependence on $R_{\rm c,ISM}$ and converges to the 
Gunn \& Gott (1972) prediction given with Eq.~\ref{striprad}. Thus our 
results give much less stripping compared to Gunn \& Gott (1972) in 
clusters with a narrow ICM peak, while converging to its predictions in 
extended ICM clusters.

The convergence of the stripping level is apparent from 
Fig.~\ref{mfinsigICM}, central panel, which depicts the final ISM mass 
as a function of the column density of the encountered ICM. Clearly, 
the higher the $\Sigma_{\rm ICM}$, the more the galaxy is stripped; 
however, the dependence is not linear. The decline in the curve 
corresponds to the increasing duration of the ICM--ISM interaction and 
to the resulting stripping of more and more strippable elements. The 
final saturation than equals the Gunn \& Gott prediction when all the 
elements with $f_{rest}<p_{ram}^{max}$ are stripped. The shape of the 
dependence is unique for the given orbit, galaxy-type, and the cluster 
potential. Then, knowing from observations the galaxy's velocity and 
mass, along with the distribution of the cluster DM and $\Sigma_{\rm 
ICM}$, one can estimate the stripping and find limits on the Gunn \& 
Gott criterion. Figure~\ref{mfinsigICM}, right panel, further shows 
the reciprocal of the ISM final mass as a function of $\Sigma_{\rm 
ICM}$ for LM-type galaxy. We try to fit the dependence with 
$M_{final}^{-1} = k \Sigma_{ICM} + 1$, where $k\sim 0.09$. The quantity 
$M_{final}^{-1}$ in fact corresponds to $M_{ISM,init}/ M_{ISM,final}= 
M_{HI,expected}/ M_{HI,observed}$, whose logarithm then gives the 
deficiency DEF introduced by Solanes et al. (1996).

\begin{figure}[t]
\centering
\includegraphics[width=0.24\textwidth,angle=0]{pocty_constSigICM_bound_BW.ps}
\includegraphics[width=0.24\textwidth,angle=0]{pocty_constSigICM_disk_BW.ps}
\caption{\small
Stripping of the LM-type galaxy in different clusters but with a 
similar amount of the encountered ICM, $\Sigma_{\rm ICM}\sim$ (5 -- 6) 
M$_\odot$/pc$^2$. Curves correspond to ($R_{\rm c,ICM}, \rho_{\rm 
0,ICM}$) pairs in units (kpc, 10$^{-3}$ cm$^{-3}$). The ISM mass bound 
to the galaxy (left) or staying within $r<16$ kpc and $|z|<1$ kpc layer 
about the disk plane (right) are displayed. 
}\label{pocty_constSigICM}
\end{figure}

\begin{figure*}[t]
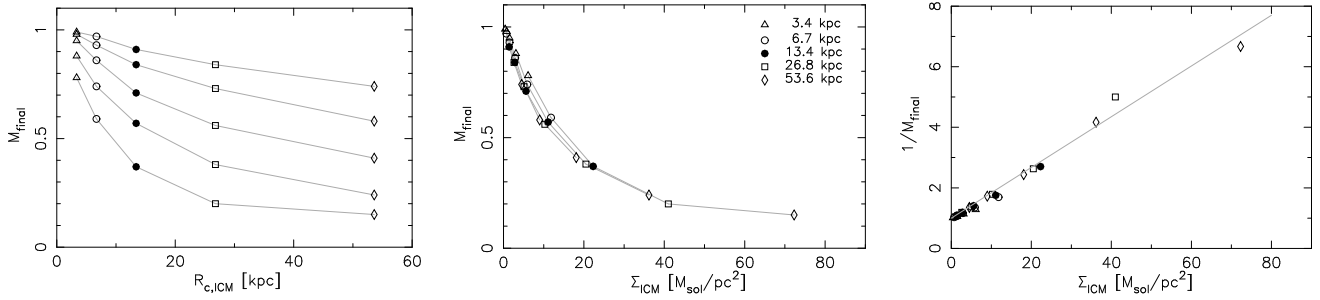

\centering
\includegraphics[width=0.3\textwidth]{finalM_Rc_paper.ps}
\hspace{0.25cm}
\includegraphics[width=0.3\textwidth]{finalM_SigICM_paper.ps}
\hspace{0.25cm}
\includegraphics[width=0.3\textwidth]{1_over_finalM_SigICM.ps}
\caption{\small
Left: Final ISM mass fraction $M_{\rm final}$ as a function of $R_{\rm 
c,ICM}$ for the LM-type galaxy. The lines correspond to different 
values of the $\rho_{\rm 0,ICM}$: 1, 2, 4, 8, and 16 $\times$ 10$^{-3}$
cm$^{-3}$ (from top down); see legend in the central panel. Center: 
$M_{\rm final}$ as a function of the column density of the encountered 
ICM, $\Sigma_{\rm ICM}$, for the LM-type galaxy. Right: Inverse value 
of $M_{\rm final}$ as a function of $\Sigma_{\rm ICM}$.
}\label{mfinsigICM}
\end{figure*}

\subsection{Various galaxy types}\label{subsec:various}
\begin{figure*}[t]
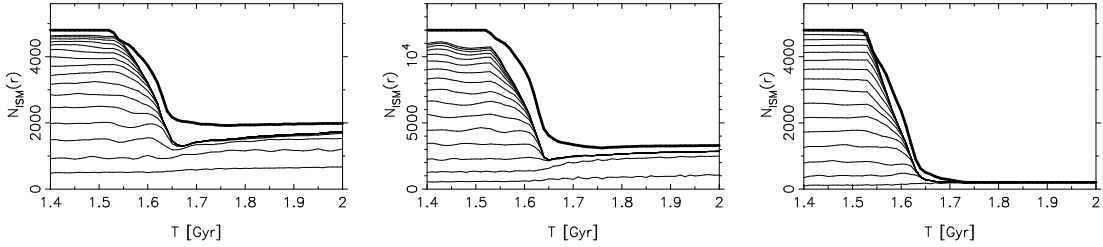

\centering
\includegraphics[height=0.25\textwidth,angle=270]{num_Lm.ps}
\hspace{0.25cm}
\includegraphics[height=0.25\textwidth,angle=270]{num_EMass.ps}
\hspace{0.25cm}
\includegraphics[height=0.25\textwidth,angle=270]{num_Em.ps}
\caption{\small
Evolution of the stripping of the Lm, EM, and Em galaxy types (from
left to right) during their crossing of the standard cluster.
}\label{LMLmEMEm_res}
\end{figure*}

A comparison of the models Lm, EM, and Em in the standard cluster is
shown in Fig. \ref{LMLmEMEm_res}. The LM-type galaxy has a stripping
radius of about 6 kpc and loses about 29\% of its original ISM. For
Lm and EM-type galaxies the stripping radii move closer to the galaxy
center, to about 3.9 kpc and 2.9 kpc, respectively. The Em-type
galaxy is stripped almost completely. As summarized in Table
\ref{tab_sim_types}, the LM, Lm, EM, and Em galaxy types lose about 
29\%, 59\%, 69\%, and 96\% of its original ISM during the face-on 
motion through the cluster with standard ICM distribution.

We may ask why the amount of the stripping is smaller in the case of
the Lm-type compared to the EM-type, when the maximum restoring force 
is stronger for the EM galaxy except the innermost radii (see 
Fig.~\ref{frestLE}, left panel). However, in the range (1 -- 4) kpc,
where the stripping radii of the two cases reside, the maxima of the
restoring forces are almost equal. Further, the surface density of the
ISM is lower, and the escape velocity is higher in the EM-type compared 
to the Lm-type. But, as we see later, the combination of a lower escape 
velocity and a higher $\Sigma_{\rm ISM}$ in the Lm-type does not 
explain less stripping compared to the EM-type.

\begin{figure*}[t]
\centering
\includegraphics[width=0.98\textwidth,angle=0]{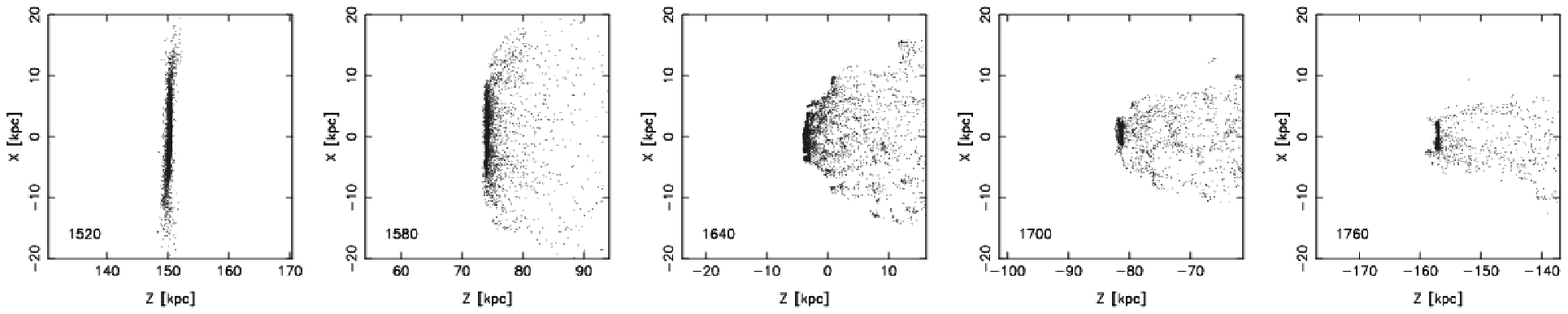}
\hspace{0.25cm}
\includegraphics[width=0.98\textwidth,angle=0]{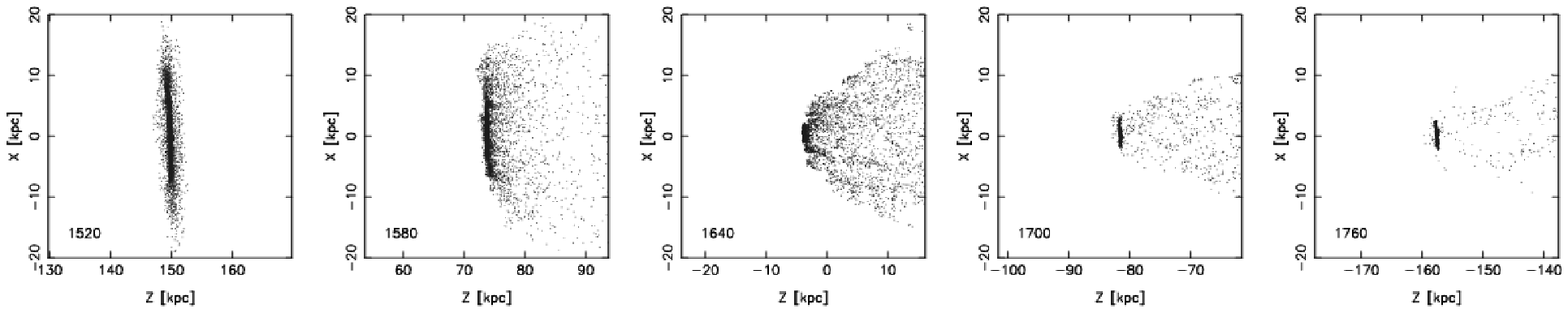}
\caption{\small
Edge-on snapshots of the stripping of the Lm (top) and EM (bottom)
galaxies. The ISM disk of the EM galaxy type is clearly thicker in
comparison with the Lm-type before they enter the ICM distribution.
}\label{Lm_vs_EM}
\end{figure*}

As shown in the middle panel of Fig.~\ref{frestLE}, the maxima of the 
restoring force over $z$-direction occur substantially closer to the 
disk plane in the case of the Lm galaxy, especially at small galactic 
radii. Thus, the rise in the restoring force behind the disk is much 
steeper in the Lm galaxy than in the EM model. In both cases, the 
galaxies move through the narrow standard cluster. Consequently, the 
ISM of the Lm-type galaxy is only shifted slightly out of the disk by 
the short ICM wind pulse. On the contrary, the ISM in the EM-type 
galaxy is more easily shifted to higher $z$'s by the same ram pressure 
pulse, and is thus accelerated towards the $v_{esc}$ more effectively.
This difference raises an interesting physical possibility: If the peak 
in the restoring force is a long way from the midplane, the ram 
pressure is able to do work on disk gas, building up its kinetic 
energy, before the gas encounters the peak of the restoring force. The 
stored kinetic energy could then be sufficient to get (or helping to do 
so) the gas out of the galaxy, even if the ram pressure alone is 
insufficient to do this.

The difference in the amount of ICM momentum delivered to the ISM also 
has hydrodynamical causes: the higher density of the ISM leads to
creation of a larger and stronger bow shock, better protecting the ISM
against the incoming ICM. That is more effectively deflected to the
sides, and a fraction of the ICM pressure is directed to the sides, 
thereby decreasing the postshock pressure. Then the ISM in Lm-type gets 
less ICM momenta compared to EM galaxy.

\begin{table}[t]
\centering
\begin{tabular}{ccccc}
\hline \hline
Gal.~type & R$_{strip}$ & M$_{strip}$ & $M_{min}$ & $M_{accr}$\\
& (kpc) & (\%) & (\%) & (\%)\\
\hline
LM & 6.3 & 29 & 49 & 22\\
Lm & 3.9 & 59 & 27 & 14\\
EM & 2.9 & 73 & 18 & 9 \\
Em & 0.2 & 96 & 4  & 0 \\
\hline
\end{tabular}
\caption{\small
Stripping results for different galaxy types in the standard
cluster.
}\label{tab_sim_types}
\end{table}

Figure~\ref{Lm_vs_EM} displays five snapshots of the Lm and EM galaxies
during their passage through the central parts of the standard cluster.
The ISM disk of the EM-type galaxy is much thicker than that of the 
Lm-type due to its higher bulge-to-disk mass ratio. This leads to 
another possible explanation of the more effective stripping of the EM 
galaxy, where its thicker disk is more vulnerable to the ram pressure.

\section{Simulations with varying SPH sizes of ICM particles}\label{sec:tests}
\begin{figure}[t]
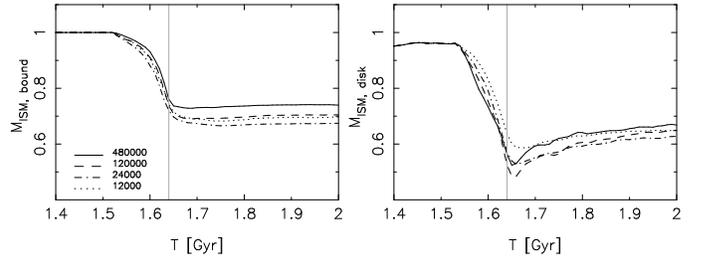

\centering
\includegraphics[width=0.24\textwidth,angle=0]{pocty_NICM_bound_BW.ps}
\includegraphics[width=0.24\textwidth,angle=0]{pocty_NICM_disk_BW.ps}
\caption{\small
Simulation results of the LM-type galaxy crossing the standard cluster 
with a varying number of ICM particles. Total mass of the bound ISM 
(left) and of the ISM located within $r<16$ kpc, $|z|<1$ kpc zone 
about the disk plane (right) are displayed as functions of the orbital 
time. The vertical line indicates the instant of the passage through the 
cluster center.
}\label{pocty_NICM}
\end{figure}

To test the influence of the number of ICM particles, we vary $N_{\rm 
ICM}$ by a multiplicative factor $\alpha$. Then, since the SPH number 
of neighbors is fixed, the smoothing lengths evolve as $h'=h\ 
\alpha^{-1/3}$. As the number of ICM particles in the simulation 
increases, the hydrodynamical interactions are treated in a better way, 
since the mutual sizes of ICM and ISM particles approach. We have
performed a set of test runs with the LM-type galaxy crossing the
standard cluster with a number of ICM particles varying as 480\,000,
120\,000, 24\,000, and 12\,000. Figure~\ref{pocty_NICM} displays the
results -- the time evolution of the ISM mass bound to the galaxy
(top), and the ISM mass located within the $r<16$ kpc, $|z|<1$ kpc
zone (bottom). Although the number of ICM particles differs by a
factor 40  in the extreme cases, which corresponds to about a 
3.4-fold difference in the SPH particle sizes, only a small
difference in the final stripping results is observed  in
Fig.~\ref{pocty_NICM}.

Since all particles meeting the condition $|${\boldmath$r$}$_{ij}|<
\max(h_i,h_j)$ are used to evaluate the pressure gradient at the disk, 
the ISM particles still feel the static pressure of the ICM behind the 
bow shock, even though the SPH sizes of the ICM particles may be much 
greater than those of the ISM. This view is supported by 
Fig.~\ref{pocty_NICM}, since changing the number of ICM particles does 
not alter the stripping rate. The dominant process governing the 
sweeping in the case of supersonic ISM--ICM interaction is the ram 
pressure.

Figure~\ref{pocty_NICM} also shows that, due to smoothing of the ICM 
peak by increasing sizes of the particles, shifting of the ISM out 
of the $|z|<1$ kpc evaluation zone is steeper. Apart from changing 
sizes of the ICM particles, the growing value of the $N_{\rm ICM}$ 
causes different mass ratios of individual ICM vs. ISM particles: 10.3, 
5.2, 1, and 0.26.

\section{Analytical model}\label{sec:analyt}
After presenting our numerical results, we now consider the interaction 
of the galactic ISM with the ICM from an analytic point of view.

\subsection{The equation of motion of an ISM element}
We describe the trajectory of an individual ISM element originally 
sitting in the disk plane, as it moves under the influence of the 
cluster gravity forces, the restoring force of the galaxy, and the ICM 
pressure force.

\begin{figure}[t]
\centering
\includegraphics[clip=true,bb=316 591 579 769,width=0.4\textwidth]{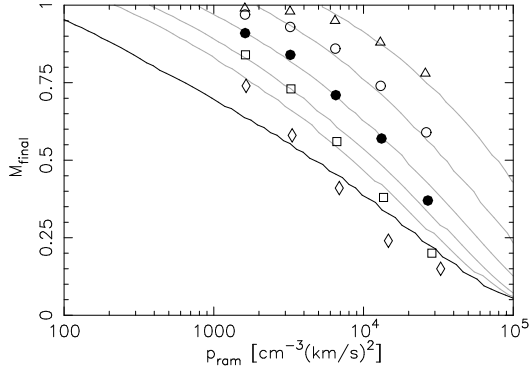}
\caption{\small
The final mass fraction $M_{\rm final}$ of the gas that remains in the 
disk after the stripping events as a function of the maximum value of 
the ram pressure $p_{ram}$. The symbols (see legend in 
Fig.~\ref{mfinsigICM}) denote the results of the SPH calculations for 
five groups of clusters with increasing central density (from left to 
right), while the gray curves are the corresponding predictions 
obtained by integration of Eq.~\ref{eom}. The thick solid line refers 
to the criterion of Gunn \& Gott (1972), Eq.~\ref{striprad}. 
}\label{eom_simul}
\end{figure}

As shown in the simulations, a bow shock forms when the incoming ICM
arrives at the galaxy's disk. Consequently, the ICM decelerates at the 
shock, transferring  part of its momentum to the galaxy, and flows 
along the shock envelope around the disk. The disk itself feels the 
pressure of the bow shock and thus the ISM is pushed out of the disk 
plane, against the galaxy's restoring force. 

Disregarding the above hydrodynamical effects and considering the 
face-on case, the equation of motion (EOM) for the movement of an ISM 
element in the $z$-direction, i.e. vertical to the galactic plane, is
\begin{equation}\label{eom}
\frac{d\,(v_{out}\Sigma_{\rm ISM})}{dt} = \rho_{\rm ICM}(v-v_{out})^2 -
\frac{\partial\Phi}{\partial z}\, \Sigma_{\rm ISM},
\end{equation}
where $v$ is the galaxy's orbital velocity. The rightmost term is the 
gravitational restoring force at the current $z$-position of the ISM 
element, whose current velocity is $v_{out}$ with respect to the 
galaxy's rest frame. The ISM element feels a reduced ram pressure 
because of the relative velocity of the galaxy, and the ICM, $v$, is 
reduced by $v_{out}$. When $v_{out}$ exceeds the local value of the 
escape velocity $v_{esc}(r,z)$, the element is stripped.

\begin{figure*}[t]
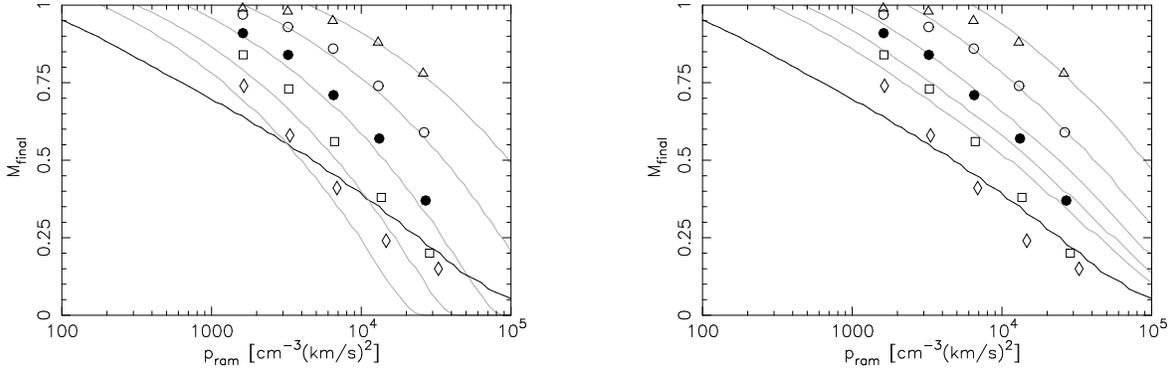

\centering
\includegraphics[clip=true,bb=316 591 579 769,width=0.4\textwidth]{formula_9.ps}
\hspace{1cm}
\includegraphics[clip=true,bb=316 591 579 769,width=0.4\textwidth]{formula_10.ps}
\caption{\small 
Numerical results versus analytic predictions of Eqs.~\ref{vafterI+} 
(left) and \ref{vafterII} (right) of final remaining gas mass fraction 
$M_{\rm final}$ after the stripping events as functions of the maximum 
value of the ram pressure $p_{ram}$. The symbols for individual 
simulations are the same as in Fig.~\ref{eom_simul}. The thick solid 
line corresponds to Gunn \& Gott (1972), Eq.~\ref{striprad}. 
}\label{formulae_9_10}
\end{figure*}

The solution of this equation, together with $dz/dt = v_{out}$, gives 
the speed and the trajectory of the ISM element in the gravitational 
field of the galaxy and under the influence of the changing ram 
pressure. From Eq.~\ref{eom} one notes that, apart from the actual 
value of the ram pressure reduced by the local restoring force, the 
history of the encountered ICM also matters. In the disk of a galaxy, 
one applies this EOM to the ISM at each galactocentric radius; and by 
numerical integration, one can follow the fate of each annulus of the 
disk until it is stripped. The stripping radius $r_{strip}$ is a radius 
above which $v_{out}(r)$ exceeds the value of the escape velocity in 
the galactic plane. The final mass is then determined as the mass of 
the ISM enclosed within the stripping radius, according to the density 
profile in the disk (Eq.~\ref{disk}). 

Figure~\ref{eom_simul} compares these predictions of Eq.~\ref{eom} with 
the results of the SPH simulations, depicting $M_{\rm final}$ as a 
function of the peak value $p_{ram}$ of the ram pressure. One notes that 
the curves from the simple EOM model match the results obtained by the 
SPH simulations rather well, in particularly for narrow clusters 
with $R_{\rm c,ICM} = 3.4$, 6.7, and 13.4~kpc. For large clusters, the 
simple model underestimates the stripping severely. This may be because 
in Eq.~\ref{eom} we neglect the deformation of the stripped ISM disk, 
which would affect the restoring force, or none of the hydrodynamical 
effects is taken into account. Nonetheless, this very simple approach
reproduces the dependence on both the extend and the density of the
cluster quite well.

Figure~\ref{eom_simul} also shows that the criterion of Gunn \& Gott 
(1972) constitutes the limit for maximum possible stripping. The final 
mass fractions predicted from the simple approach converge towards the 
estimates based on Eq.~\ref{striprad} at high values of $R_{\rm 
c,ICM}$ and $\rho_{\rm 0,ICM}$, but they always remain higher. This is 
because Gunn \& Gott (1972) assume a constant operation of the peak 
value of the ram pressure and compare it with the maximum restoring 
force. Therefore, at all radii of the disk, where the (peak) ram 
pressure exceeds the maximum restoring force, all the ISM elements can 
be accelerated to the escape velocities. However, in the integration of 
the EOM, an ISM element may cross the $z$-distance of the maximum 
restoring force, but the peak of the ram pressure may be over before 
the element reaches escape velocity.

\subsection{The impulse approximation}
Since our principal intention is not to follow the exact trajectory of 
the stripped elements but only to identify them and thus obtain the 
stripping radius (or the final mass), let us introduce an approximative 
solution of the EOM.

If we consider only those regions where the gas is clearly removed from 
the galaxy disk, we can obtain a first-order solution by considering 
the balance of the momentum. Neglecting the restoring force in these 
outer regions, the momentum per unit surface area attained by the 
stripped gas of surface density $\Sigma_{\rm ISM}$ and velocity 
$v_{after}$ is 
\begin{equation}\label{vafterI}
P_{after} = v_{after} \Sigma_{\rm ISM}= \int \rho_{\rm ICM} v^2 dt.
\end{equation}
This quantity $P_{after}$ can be considered as the integrated history 
of the ram pressure:
\begin{equation}\label{pafterII}
P_{after} = \Sigma_{\rm ICM} {\int \rho_{\rm ICM} v^2 dt \over \int 
\rho_{\rm ICM} v dt} = \Sigma_{\rm ICM} \langle v\rangle_{\rho_{\rm 
ICM}},
\end{equation}
where $\Sigma_{\rm ICM}=\int \rho_{\rm ICM} v dt$ is the column density 
of the encountered ICM, and $\langle v\rangle_{\rho_{\rm ICM}}$ is the 
averaged velocity $v$ along the galaxy's orbit weighted with the local 
volume density $\rho_{\rm ICM}$. Then Eq.~\ref{vafterI} becomes
\begin{equation}\label{vafterI+}
v_{after} = \langle v\rangle_{\rho_{\rm ICM}} {\Sigma_{\rm ICM} \over
\Sigma_{\rm ISM}}.
\end{equation}

Of course, the restoring force $f_{rest}$ can only be neglected in the 
stripped outer parts of the galaxy. A better approximation of the EOM 
is to consider the momentum transferred from the inflowing ICM to the 
ISM element after the element passed the position $z$ with maximum 
value of the restoring force. Thus, one regards the acceleration of the 
ISM only when the ram pressure exceeds this maximum. Then, at a given 
radius, 
\begin{equation}\label{vafterII}
v_{after} = \frac{\int (\rho_{\rm ICM} v^2 - {\partial \Phi \over 
\partial z}|_{\rm max}\, \Sigma_{\rm ISM})\, dt}{\Sigma_{\rm ISM}},
\end{equation}
where the integration is taken over a time interval ($t_1,t_2$) when 
$p_{ram}>f_{rest,max}$, i.e. as long as the Gunn \& Gott (1972) 
criterion (Eq.~\ref{striprad}) is fulfilled. Since the actual value of 
the restoring force at the position $(r,z)$ is less compared to its 
maximum value, this obviously overestimates the binding of an element 
to the galaxy plane. 

Figure~\ref{formulae_9_10} shows a comparison of the results of the SPH 
simulations and from the approximative solutions by 
Eqs.~\ref{vafterI+} and \ref{vafterII}, which are expected to bracket 
the actual equation. As before, the stripping radius $r_{strip}$ is the 
radius above which $v_{after}(r)$ exceeds the value of the escape 
velocity in the $z=0$ plane. Then the final mass is determined as the 
mass of the ISM enclosed within the stripping radius, according to the 
disk's density profile (Eq.~\ref{disk}). One notes that the curves 
from Eq.~\ref{vafterI+}, which neglects the restoring force, indeed 
overestimate the stripping, while curves based on Eq.~\ref{vafterII}, 
which consider only the momentum gained as long as the ram pressure 
exceeds the maximum restoring force, underestimate the stripping. In 
any case, the departures from SPH results are larger than with the 
predictions of the EOM (Eq.~\ref{eom}), displayed in 
Fig.~\ref{eom_simul}.

Of course, comparing the full numerical treatment with the predictions 
of the analytical approach may have several caveats. The most important 
is that all the hydrodynamical processes, such as the effect of the 
bow-shock and size of the galaxy shadow, are disregarded in the 
solution that describes the dynamics of a single ISM element. 
Furthermore, in the analytic calculations, the density profile of the 
ISM is assumed to be constant, whereas in the full simulations, 
processes of relaxation and formation of spiral arms, as well as the 
progressive stripping, can change the surface density profile of the 
gas disk. Such changes consequently affect the restoring force and the 
escape velocity profiles. 

Nonetheless, it is interesting that the simple conservation of momentum 
(Eq.~\ref{vafterI+}), which on one hand neglects the actual effect of 
the restoring force and the element's shifting out of the disk and on 
the other overestimates the escape velocity and changes of the 
ISM disk due to stripping, can give a good order of magnitude estimate 
of the stripping, when the variation in the ICM density along the 
galaxy's path is to be taken into account. This corresponds to the 
above fit of simulation results with $M_{final} \sim (k \Sigma_{ICM} + 
1)^{-1}$: for higher $\Sigma_{ICM}$, $v_{after}$ grows, turning 
more and more elements to higher than escape velocity $v_{esc}$. 
Therefore $M_{final}$ decreases with growing $\Sigma_{ICM}$.
    
The Gunn \& Gott (1972) formula, computed from the maximum ICM density, 
overestimates the stripping and does not allow for the effect of the 
finite size of the cluster, i.e. the finite time of interaction. 
However, in cases with high values of $R_{\rm c,ICM}$ and $\rho_{\rm 
0,ICM}$, the SPH simulations show an even larger stripping than the Gunn 
\& Gott (1972) estimate. This may be a consequence of uneven gas 
distribution in the galaxy disk where spiral-arm inhomogeneities 
contribute to the ISM removal, as also shown by Quilis et al.~(2000).

\section{Comparison to other simulations}\label{sec:comparison}
\subsection{Abadi et al.~(1999)}
Performing 3D SPH/N-body simulations, Abadi et al.~(1999) studied the 
stripping of spiral galaxies by operating a constant flow of ICM 
particles in a simulation box of size 60 kpc $\times$ 60 kpc $\times$ 
10 kpc homogeneously filled up with the ICM particles flowing with 
periodic boundary conditions. Their model is a spiral galaxy exposed to 
a wind of 1000 km s$^{-1}$, 2000 km s$^{-1}$, or 3000 km s$^{-1}$ and of 
density: $\rho_c=3.37\ 10^{-3}$ cm$^{-3}$, and $\rho_v = 0.1\rho_c$. 
The ram pressure then ranges from about 300 to 30\,000 cm$^{-3}$ km$^2$ 
s$^{-2}$. For the higher value of the ICM density, they estimate the 
stripping radius of a spiral galaxy to 4 kpc, i.e. to about 80~\% loss 
of the galaxy's diffuse gas mass. 

\subsection{Quilis et al.~(2000)}
In their simulations employing a high-resolution 3D Eulerian code with 
a fixed grid based on high-resolution shock-capturing method, Quilis 
et al. (2000) follow the interaction between the hot ICM and the cold 
ISM. The stellar and DM components are evolved using a particle-mesh 
code. They model a luminous spiral galaxy similar to the Milky Way or 
Andromeda. The ICM is modeled as a uniform medium with temperature 
$T_{\rm ICM}=10^8$ K and constant densities $\rho_c=2.6\ 10^{-3}$ 
cm$^{-3}$ and $\rho_v = 0.1\rho_c$. The galaxy moves through the ICM 
wind at velocities of 1000 km s$^{-1}$ or 2000 km s$^{-1}$. For the 
Coma-like ICM density and velocity, they estimate the stripping radius 
as $\sim 3$ kpc. 

\subsection{Vollmer et al.~(2001)}\label{sec:Vollmer2001}
Investigating the role of the ram pressure stripping in the Virgo 
cluster, Vollmer et al.~(2001) employ a method of sticky particles for 
modeling the warm neutral clouds of the ISM. The effect of the ram 
pressure is included only analytically as an additional acceleration on 
the clouds located at the windward side of the gas distribution. 
Contrary to other simulations discussed in this section, they introduce 
temporal ram pressure profiles in their simulations instead of a 
constant ICM wind. They use a spiral galaxy model in a wind creating 
the ram pressure reaching maximum values ranging from 1\,000 to 10\,000 
cm$^{-3}$ km$^2$ s$^{-2}$. 
Galaxies are considered on slightly elliptical orbits not reaching the 
very center of the cluster. Due to a high slope of the 
$\beta$-profile, the ICM density falls steeply with distance $R$ from 
the cluster center. At the smallest pericenter used ($R \sim 70$ 
kpc), the ICM density exceeds about 3 10\(^{-3}\) cm$^{-3}$. In 
our present simulations, galaxies on completely radial orbits are 
studied. To achieve similar effects of the ram pressure as Vollmer et 
al.~(2001), we have set $\rho_{\rm 0,ICM} = 0.1\times \rho_{\rm 0,ICM} 
^{\rm Vollmer}$ as our standard value (see Sect.~\ref{sec:cluster}). 

Vollmer et al.~(2001) show that the ram pressure can lead to a 
temporary increase in the central gas surface density and that in some 
cases a strong reaccretion of the atomic gas occurs after the stripping 
event. They find that the scenario of ram pressure stripping being 
responsible for the observed HI deficiency is consistent with all HI 21 
cm observations in the Virgo cluster.

The difference between a slightly elliptical orbit with the pericenter 
distance $R_p$ and a completely radial orbit with the same maximum 
value of the ICM density encountered, i.e. \(\rho_{\rm 0,ICM}^{rad} = 
\rho_{\rm ICM}^{ellipt}(R_p)\), and the same $R_{\rm c,ICM}$, is in the 
effective size of the interacting region. The width of the ram pressure 
peak is larger in the former case than in the latter one. From that it
follows that in our simulations the ISM--ICM interactions take place for 
a shorter time as compared to Vollmer et al.~(2001).

\subsection{Schulz \& Struck (2001)}
A model dwarf galaxy in 3D HYDRA (SPH-AP$^3$M) simulations of Schulz \& 
Struck (2001) is placed in a cubical grid with 100 kpc edge. The galaxy 
is surrounded with 80\,000 uniformly distributed ICM particles. The 
temperature of the ICM is $T_{\rm ICM}=4.6\ 10^5$ K, and the density of 
the wind is $\sim 7.3\ 10^{-5}$ cm$^{-3}$. In their face-on models, the 
galaxy moves with a velocity of 1000 km s$^{-1}$ or 2000 km s$^{-1}$. 
Then, the stripping radii measured from their Figs.~2 and 14 are $\sim 
7.2$ kpc and $\sim 5.2$ kpc.

\subsection{Roediger \& Hensler (2005)}
With a spiral galaxy models and a wide range of ICM conditions, 
Roediger \& Hensler (2005) perform a parameter study of the ram 
pressure stripping using 2D Eulerian simulations. Except for the 
gaseous disk, all components are treated analytically, providing only 
the gravitational potential. The ICM densities range from $10^{-5}$ to 
$10^{-3}$ cm$^{-3}$, and the resulting ram pressure covers a range from
10 to 10\,000 cm$^{-3}$ km$^2$ s$^{-2}$.\\

\begin{figure}[t]
\centering
\includegraphics[width=0.4\textwidth,angle=0]{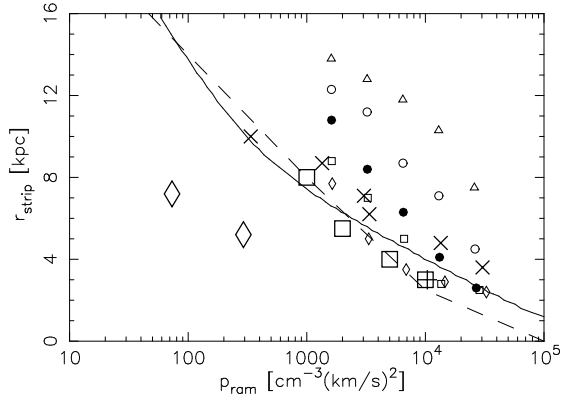}
\caption{\small
Stripping radius $r_{strip}$ versus the peak ram pressure 
$p_{ram}^{max}$: a comparison with simulations by Abadi et al.~(1999) 
-- large X; Quilis et al.~(2000) -- large plus; Vollmer et al.~(2001) 
-- large squares; Schulz \& Struck (2001) -- large diamonds; models of 
Roediger \& Hensler (2005) -- dashed line. Our simulations: $R_{\rm 
c,ICM}=$ 3.4 kpc -- triangles; 6.7 kpc -- circles; 13.4 kpc -- filled 
circles; 26.8 kpc -- squares; 53.6 kpc -- diamonds. Solid line shows 
the prediction using Gunn \& Gott (1972) (Eq.~\ref{striprad}).
}\label{other_simul}
\end{figure}

\subsection{Summary}
The different simulations are listed in Table~\ref{difsim} and a 
comparison is shown in Fig.~\ref{other_simul}. All the models shown, 
except Schulz \& Struck (2001), use comparable spiral galaxy models.
Abadi et al. (1999), Quilis et al. (2000) and Roediger \& Hensler 
(2005) show high stripping efficiency close to the Gunn \& Gott (1972) 
prediction, which is due to a constant ICM density adopted. There is no 
difference in the stripping radius and $M_{final}$ from 2D and 3D 
hydrodynamical simulations and SPH. It seems that the global stripping 
effect on the galaxy does not depend on the details of the 
hydrodynamical approach.   

\begin{table}[t]
\centering
\begin{tabular}{cccc}
\hline
\hline
Author          & Method          & ICM flow & galaxy\\
\hline
Abadi et al.    & SPH             & constant & spiral\\
Quilis et al.   & 3D finite diff. & constant & spiral\\
Vollmer et al.  & sticky particle & changing & spiral\\
Schulz et al.   & SPH-AP$^3$M     & constant & dwarf\\
Roediger et al. & 2D finite diff. & constant & spiral\\
J\' achym et al.& SPH             & changing & spiral\\
\hline
\end{tabular}
\caption{The list of simulations.}\label{difsim}
\end{table}

The results of Schulz \& Struck (2001) show the same stripping 
efficiency at lower ram pressures (see Fig.~\ref{other_simul}), which 
is, in accordance to Gunn \& Gott(1972), due to lower disk restoring 
force in the case of dwarf galaxies.

Our models with low values of $R_{\rm c,ICM}$ show less stripping. 
This is mainly due to the effect of changing strength and finite 
duration of the ram pressure peak. For high values of $R_{\rm c,ICM}$ 
in large clusters with long interaction times, our models approach the 
Gunn \& Gott (1972) prediction and the results of other models. It 
demonstrates that our SPH approach gives similar results to SPH 
simulations of others and to 2D and 3D Eulerian codes.   

Simulations by Vollmer et al. (2001) use the changing ram pressure 
force acting on sticky particles to show results close to Gunn \& Gott 
(1972) prediction. However, with the similar dependence of the ram 
pressure on time, we get less stripping. Simulations with sticky 
particles probably overestimate the stripping efficiency, since this 
kind of simulation disregards hydrodynamical effects such as formation 
of the bow shock in the ICM in front of the galaxy. It actually shields 
the galaxy, thereby reducing the efficiency of stripping.

\section{Discussion and conclusions}
The simulations performed in the present work allow us to estimate the 
consequences of ram pressure stripping in clusters better. If a galaxy 
like the Milky Way (our type LM) passes in face-on orientation through 
the central parts of a cluster similar to the Virgo, it loses about 
30\% of its interstellar gas. This constitutes a rather large amount of 
gas loss. According to Roediger \& Br\"uggen 
(2006), the efficiency of stripping only marginally depends on the 
orientation of the galaxy for inclinations lower than $\sim 60^\circ$. 
Therefore, we disregard this aspect in this discussion.

With this level of stripping, we may estimate the fraction of gas 
supplied to the ICM by stripping. About 2 10$^9$ M$_\odot$ of gas is 
lost by a LM-type galaxy. A much less massive Lm-type galaxy 
is more severely stripped, thus roughly providing the same amount of 
gas. EM and Em-types are subject to an even more effective stripping, 
but their gas disks are less massive. Our estimate is that the average 
contribution per spiral galaxy, late or early type, with different 
orientations in a radial orbit, is 1 -- 2 10$^9$ M$_\odot$. This means 
that the total amount of hot gas in the central region of our standard 
cluster (6 10$^{10}$ M$_\odot$) can be provided by 30 -- 60 stripping 
acts. The amount of stripping from past to present is influenced with 
the decreasing gas content of galaxies, which is an obvious consequence 
of gas consumption by star formation: gas-rich galaxies in the past can 
provide more gas to intergalactic space. In contrast, the growing 
ICM concentration in the center of the cluster leads to higher 
stripping efficiency now.

Note that only the contribution from galaxies on radial orbits is 
substantial. Galaxies on circular or elliptical orbits do not penetrate 
the dense parts of the ICM and thus are safe from being strongly 
stripped. Depending on the fraction of galaxies in radial orbits in the 
cluster, which unfortunately is not well known, the stripping could 
supply a significant amount of gas to the ICM. This may explain the 
high metallicity of the ICM gas.

Another consequence of our simulations is that we expect to find a 
significant amount of relatively cold gas in any cluster, as debris left 
over from recent stripping events. This gas is not yet mixed with the 
ICM, but forms large diffuse clouds as tails behind the galaxy keeping 
a fraction of its velocity. The mixing time should be comparable to the 
free-fall time, e.g. $\sim $ 1 Gyr. The density of gas in tails may be 
more than 10 times higher than the local hot ICM density, hence a tail 
could cause strong stripping for any galaxy that happens to cross it 
at a very different velocity. This can then provide a nice explanation 
of the puzzling strong stripping recently observed in many regions 
where the hot ICM gas is not sufficient to strip. For instance NGC 4522 
is apparently stripped at a large distance from the cluster center 
(Crowl \& Kenney 2006), where the ICM density is not sufficient. A nice 
example of a cold HI tail almost 100 kpc long provides the galaxy CGCG 
97-073 in the Coma cluster. This and similar tails may be the cause of 
ram pressure for other galaxies arriving at high speed. The 
surroundings of Virgo M86 galaxy shows a complex structure in X-ray 
emission (Finoguenov et al. 2004), like a large plume extending in the 
northwest direction. However, the influence of such spatially 
narrow and small ICM peaks acting on a galaxy during a short time is 
estimated to be very slow. 
Also, ram-pressure stripping appears in small groups, where the X-ray 
gas is not detected (Rasmussen et al. 2006). As we have shown, the 
crucial effect of the ram pressure stripping is the pressure itself, so
the hydrodynamical effects play only a minor role. Therefore, not only 
is the hot ICM able to strip the galactic ISM, but debris from tidal 
interactions or previous stripping do so, as well.

The main conclusion is that the stripping efficiency depends 
significantly on the duration of the ram pressure pulse, which in many 
cases means that the Gunn \& Gott (1972) prediction overestimates the 
amount of stripping. This is nicely shown by Boselli \& Gavazzi (2006, 
their Fig.~18), where they compare the restoring force in spiral 
galaxies in Coma, A1367 and Virgo clusters to the maximum ram pressure. 
It turns out that many galaxies with normal HI content should have been 
stripped if the ram pressure had acted long enough. 
Much more modest stripping is observed, showing that Gunn \& Gott (1972) 
prediction is an overestimate, since it neglects the role of 
reaccretion of the ISM. 
As encountered in Sect.~\ref{subsec:various}, in early-type 
galaxies with high bulge-to-disk mass ratios, the maximum of the 
gravitational restoring force is more displaced from the disk plane 
than in late types. This means that the ram pressure can accelerate the 
disk gas more before it reaches the peak of the restoring 
force. In principle, this mechanism could allow stripping even when the 
Gunn \& Gott condition is not met. The right panel of Fig.~\ref{frest} 
shows that, especially at larger disk radii, the maxima of the restoring 
force are well off the galaxy midplane. Our simulation results support 
this idea -- in Fig.~\ref{other_simul} galaxies crossing clusters with 
large ICM distributions show more stripping than corresponds to the 
Gunn \& Gott prediction.

\section{Summary}
Using the SPH/tree code GADGET (Springel et al. 2001) adapted for 
interactions of two gaseous phases, the ISM and the ICM, we have 
performed simulations of the ram pressure stripping of galaxies 
on completely radial orbits crossing galaxy clusters with realistic 
profiles of the ICM density. 

We found different amounts of stripping (from complete to only 
marginal) in clusters with various ICM distributions. However, when 
galaxies encounter the same amount of ICM on their orbits, the 
stripping results are found to be the same. Thus, for a given orbital 
type and cluster potential, the column density $\Sigma_{\rm ICM}$ of the 
crossed ICM determines the stripping (see Fig.~\ref{mfinsigICM}). We 
have proposed a simple analytic stripping criterion 
(Eq.~\ref{vafterI+}) taking the $\Sigma_{\rm ICM}$ into account.

Since simulations with different levels of hydrodynamical treatment show 
similar results, we have suggested that the ram pressure is the dominant 
process in sweeping of galaxies. Although this also reflects 
the formation of a bow-shock structure in front of the galaxy due to 
its supersonic motion, we proposed the important role played by the ram 
pressure in the interactions of galaxies with debris structures in 
clusters.

We have indicated the limits of the Gunn \& Gott (1972) stripping 
formula, on one hand overestimating the amount of stripping since 
completely neglecting the time-dependence of the ISM--ICM interaction 
process, and on the other underestimating the stripping by omitting the 
accumulation of kinetic energy by the gas before it encounters the 
maximum of the restoring force.

\begin{acknowledgements}
The authors gratefully acknowledge support by the Institutional 
Research Plan AV0Z10030501 of the Academy of Sciences of the Czech 
Republic and by the project LC06014 Center for Theoretical 
Astrophysics. The majority of the simulations were carried out on the 
IBM Power 4 processors of the CNRS computing center at the IDRIS 
(Palaiseau, France). We would like to thank the anonymous referee for 
helping us to improve this paper considerably.
\end{acknowledgements}

%~\newpage
%~\newpage

\begin{figure*}[t]
\centering
\includegraphics[height=0.98\textwidth,angle=270]{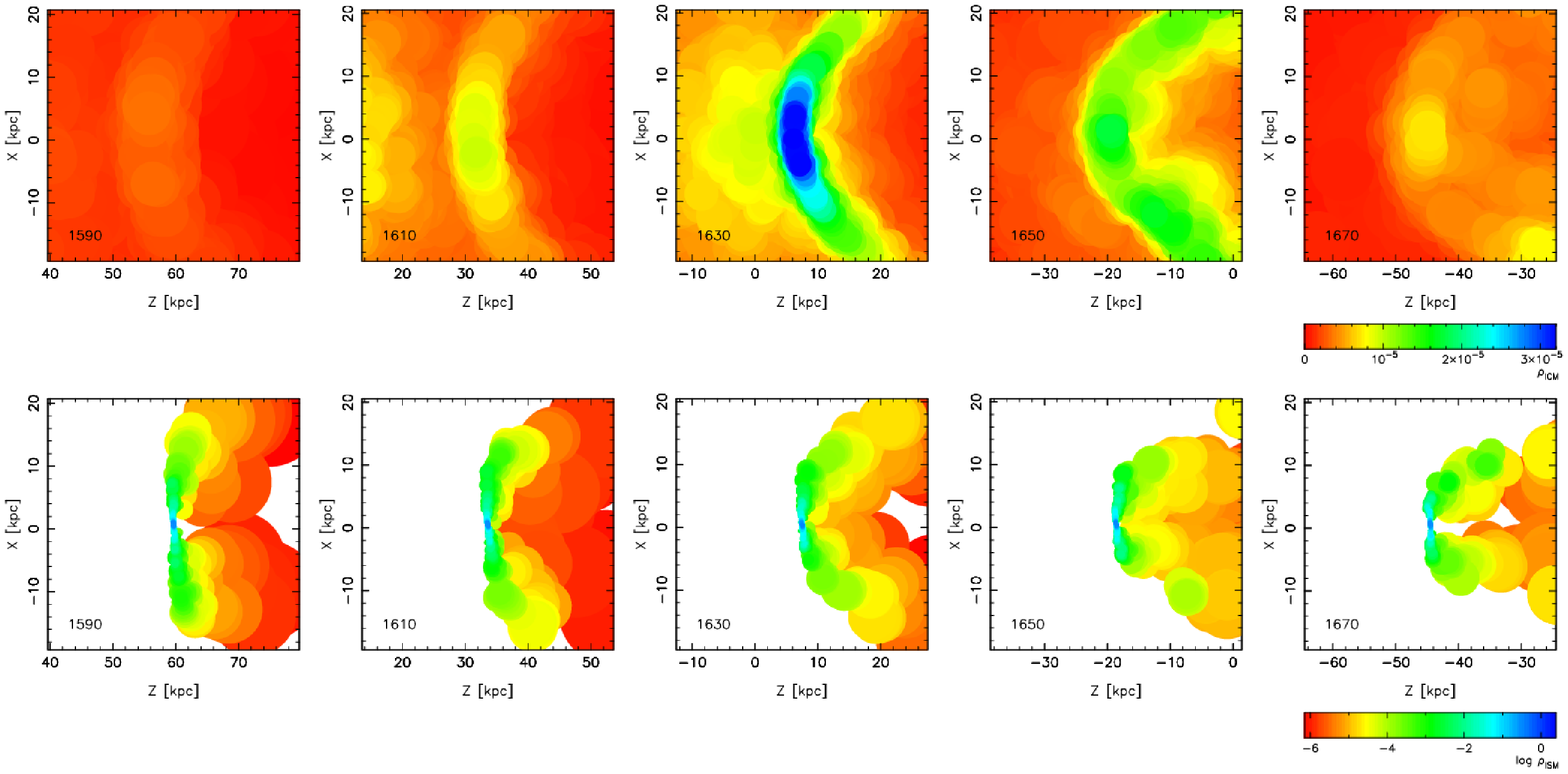}
\caption{\small Density of the ICM in surroundings of the LM-type 
galaxy (top) and of the ISM (bottom). $N_{\rm ICM}=480\,000$. Particles are 
displayed as filled circles with radii equal to their SPH smoothing 
sizes. The ISM density scale is logarithmic. The time in Myr is given 
in the left lower corner of frames. 
}\label{den480}
\end{figure*}

\begin{figure*}[t]
\centering
\includegraphics[height=0.98\textwidth,angle=270]{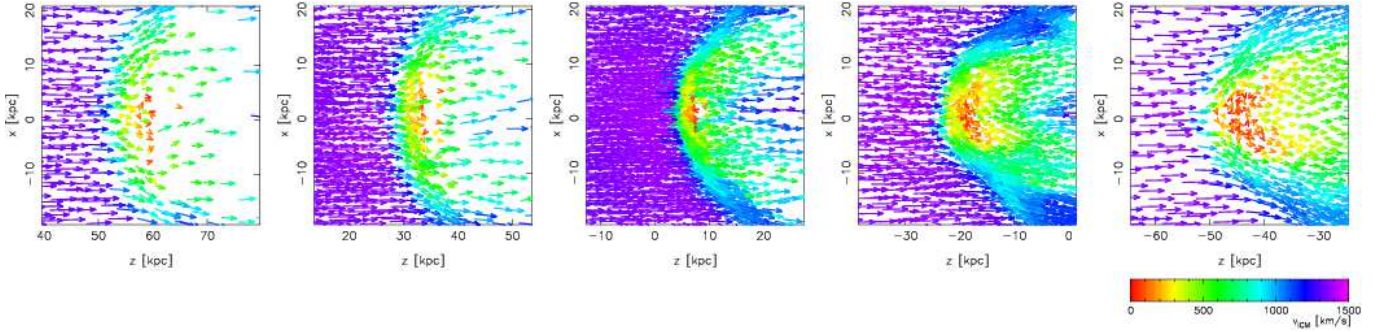}
\caption{\small
The flow of ICM particles around the LM-type galaxy in the case of 
$N_{\rm ICM}=480\,000$.
}\label{velo480}
\end{figure*}

~\newpage
~\newpage
\section*{Appendix}
In Sect.~\ref{LMinstand}, Figs.~\ref{den} and \ref{velo} show the 
distribution of the ICM density and relative velocity at five locations 
of the LM galaxy when it crosses the standard cluster. There, the ICM is 
represented with 120 000 SPH particles. Figures~\ref{den480} and 
\ref{velo480} depict the same situation with $N_{ICM}=480 000$. Now, 
ICM particles are smaller and the bow shock in front of the galaxy is 
more pronounced and ISM in the disk slightly better protected. 
However, as shown in Sect.~\ref{sec:tests}, the stripping results are 
almost independent of the value of $N_{ICM}$.

\end{document}